\documentclass[iop]{emulateapj}
\usepackage{color}

\usepackage{enumitem} 

\newcommand{\hMsun}{{\ifmmode{h^{-1}{\rm
        {M_{\odot}}}}\else{$h^{-1}{\rm{M_{\odot}}}$~}\fi}} 
\newcommand{\hMpc}{{\ifmmode{h^{-1}{\rm Mpc}}\else{$h^{-1}$Mpc }\fi}}
\def\be{\begin{equation}}
\def\ee{\end{equation}}
\def\ba{\begin{eqnarray}}
\def\ea{\end{eqnarray}}

\shorttitle{Reshift Dependent Volume effect}
\shortauthors{X.-D. Li, C. Park, C.G. Sabiu, H. Park, C. Cheng, J. Kim, S.E. Hong}

\begin{document}


\title{Cosmological constraints from the redshift dependence of the volume effect using the galaxy 2-point correlation function across the line-of-sight}



\author{Xiao-Dong Li, Changbom Park, }
\affil{School of Physics, Korea Institute for Advanced Study, 85 Heogiro, Dongdaemun-gu, Seoul 130-722, Korea}
\author{Cristiano G. Sabiu\altaffilmark{1}, Hyunbae Park, }
\affil{Korea Astronomy and Space Science Institute, Daejeon 305-348, Korea}
\author{Cheng Cheng, }
\affil{Kavli Institute for Theoretical Physics China, Institute of Theoretical Physics, Chinese Academy of Sciences, Zhong Guan Cun Street 55\#, Beijing, 100190, P.R. China}
\affil{University of Chinese Academy of Sciences, P.R. China}
\author{Juhan Kim,}
\affil{Center for Advanced Computation, Korea Institute for Advanced Study, 85 Hoegi-ro, Dongdaemun-gu, Seoul 130-722, Korea}
\and
\author{Sungwook E. Hong}
\affil{School of Physics, Korea Institute for Advanced Study, 85 Heogiro, Dongdaemun-gu, Seoul 130-722, Korea}
\affil{Korea Astronomy and Space Science Institute, Daejeon 305-348, Korea}

\altaffiltext{1}{Corresponding Author: csabiu@kasi.re.kr}


%
%
%
%
%


\begin{abstract}
We develop a methodology to use the redshift dependence of the galaxy 2-point correlation function (2pCF) across the line-of-sight, $\xi(r_{\bot})$, as a probe of cosmological parameters. 
The positions of galaxies in comoving Cartesian space varies under different cosmological parameter choices, 
inducing a {\it redshift-dependent scaling} in the galaxy distribution. 
This geometrical distortion can be observed as a redshift-dependent rescaling in the measured $\xi(r_{\bot})$. 
We test this methodology using a sample of 1.75 billion mock galaxies at redshifts 0, 0.5, 1, 1.5, 2, 
drawn from the Horizon Run 4 N-body simulation.
The shape of $\xi(r_{\bot})$ can exhibit a significant redshift evolution when the galaxy sample is 
analyzed under a cosmology differing from the true, simulated one. 
Other contributions, including the gravitational growth of structure, 
galaxy bias, and the redshift space distortions, 
do not produce large redshift evolution in the shape. 
We show that one can make use of this geometrical distortion to constrain the values of 
cosmological parameters governing the expansion history of the universe. 
This method could be applicable to future large scale structure surveys, 
especially photometric surveys such as DES, LSST, to derive tight cosmological constraints. 
This work is a continuation of our previous works \citep{Li2014,Li2015,Li2016} 
as a strategy to constrain cosmological parameters using redshift-invariant physical quantities.
\end{abstract}





\section{Introduction}

The discovery of cosmic acceleration \citep{Riess1998,Perl1999} implies either the existence of a ``dark energy'' component in the Universe 
or the breakdown of Einstein's gravity theory on cosmological scales \citep[see ][for a recent review]{2012IJMPD..2130002Y}.
The theoretical explanation and observational probes of cosmic acceleration have attracted 
tremendous attention in the last two decades and are still far from being well understood or accurately measured \citep{SW1989,Li2011,DHW2013}.

In an effort to probe the cosmic expansion history, large scale structure (LSS) surveys are utilized to  measure two key geometrical quantities; 
the angular diameter distance $D_A$ and the Hubble factor, $H$. 
If they were precisely measured as a function of redshift, 
then tight constraints can be placed on cosmological parameters, 
e.g. the matter density $\Omega_m$ and the equation of state (EoS) of dark energy $w$.

When an incorrect cosmological model is assumed for transforming galaxy redshifts to comoving distances,
geometric distortions are induced in the resultant galaxy distribution. 
It should be noted that this effect is not related to the peculiar velocities of galaxies, which cause redshift-space distortions (RSD).
The distortions due to coordinate transforms include the ``volume effect'' which is the mis-scaling of the size of structures, 
and the Alcock-Paczynski (AP) effect, the shape distortion induced by the fact that distances along and perpendicular to the line of sight are fundamentally different.

Various statistical methods have been proposed to measure these effects.
The AP effect was first suggested to be measured through the anisotropic clustering of galaxies \citep{Ballinger1996,Matsubara1996}. 
Later, with the advent of large observational spectroscopic programs, this was applied to a series of LSS surveys
\citep{Outram2004,Blake2011,ChuangWang2012,Reid2012,Beutler2013,Linder2013,2014arXiv1407.2257S, Jeong2014,Sutter2014,2014ApJ...781...96L,Alam2016, Beutler2016, Sanchez2016}, 
while considering the detailed effect of systematics for unbiased estimates of $D_A(z)$ and $H(z)$ \citep{2014MNRAS.445....2V,2016MNRAS.tmp.1473R,2016arXiv160302389S}.
Alternative approaches include utilizing the AP effect using the symmetry properties of galaxy pairs \citep{Marinoni2010,Jennings2011,BB2012}
and cosmic voids \citep{Ryden1995,LavausWandelt1995,Sutter2014,Qingqing2016}.

There have also been studies proposing to measure the volume effect of this geometrical rescaling, 
these include the number counting of galaxy clusters \citep{PS1974,VL1996}, 
the topology of LSS \citep{topology}, the BAO scale \citep{EHT1998,BG03,SE03}, 
and the shape of 2pCF and power spectrum \citep{Sanchez2006,Sanchez2009}.

In \cite{Li2014,Li2015} we developed a novel strategy to probe $D_A$ and $H$ from the LSS data. 
We found that the AP effect introduces a geometric distortion that evolves significantly with redshift;
this phenomenon can be utilized to distinguish the AP effect from the effect of RSD.

The radial distances of galaxies are inferred from their measured redshifts, 
which are distorted due to the peculiar motion of galaxies. 
This leads to an apparent distortion in the redshift-space galaxy distribution \citep{FOG,Kaiser1987,Ballinger1996}. 
This RSD effect is the main systematic limiting our ability to probe the geometry of LSS from the galaxy distribution. 
However, we found that the influence of RSD is significantly reduced when focusing on the redshift dependence of galaxy clustering. 

In \cite{Li2016}, we analyzed the two-point correlation function (2pCF) of galaxies measured in the Baryon Oscillation Spectroscopic Survey (BOSS). 
We showed that the redshift dependence of the AP effect leads to tight cosmological constraints that are competitive with mainstream cosmological probes such as type Ia supernovae (SNIa), baryon acoustic oscillations (BAO), and cosmic microwave background (CMB).

In this paper we continue our exploration of redshift invariant galaxy clustering 
and present a new method for constraining cosmology using the redshift evolution of 
the galaxy two-point correlation function (2pCF) across the line-of-sight (LOS), $\xi(r_{\bot})$. 
An incorrect choice of cosmology results in {\it redshift-dependent scaling} of the galaxy distribution. 
This, in turn, results in a significant redshift evolution in the shape of $\xi(r_{\bot})$.
Of course the correlation function has other contributions factors that may induce redshift evolution, 
such as the gravitational growth of structure, galaxy bias, and redshift space distortions. 
However we find that these other contributions do not produce large redshift evolution in the shape of $\xi(r_{\bot})$. 
We test this methodology using a sample of 1.75 billion mock galaxies at redshifts 0, 0.5, 1, 1.5, 2, drawn from the Horizon Run (HR) 4 N-body simulation.

The outline of this paper proceeds as follows. 
In \S 2 we briefly review the nature and consequences of the LSS geometric distortion when performing coordinate transforms in a cosmological context. 
In \S 3 we describe the N-body mock galaxies used in the analysis.
The methodology is presented in \S 4 and \S 5.
We conclude in \S 6.

\section{Volume Effect in a Nutshell}
\label{sec:Voleffect}

\begin{figure*}
   \centering{
   \includegraphics[height=8cm]{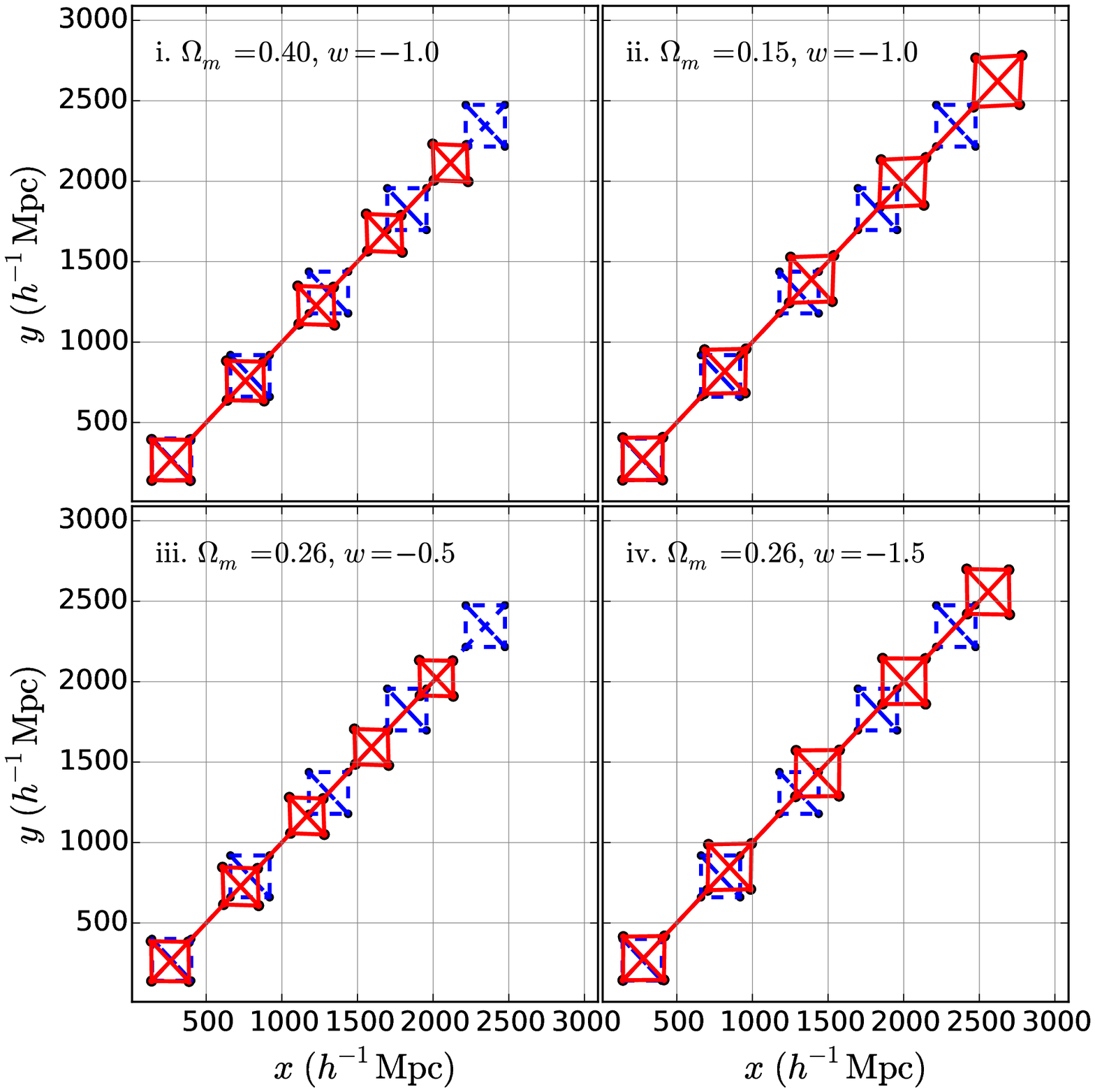}
   \includegraphics[height=8.5cm]{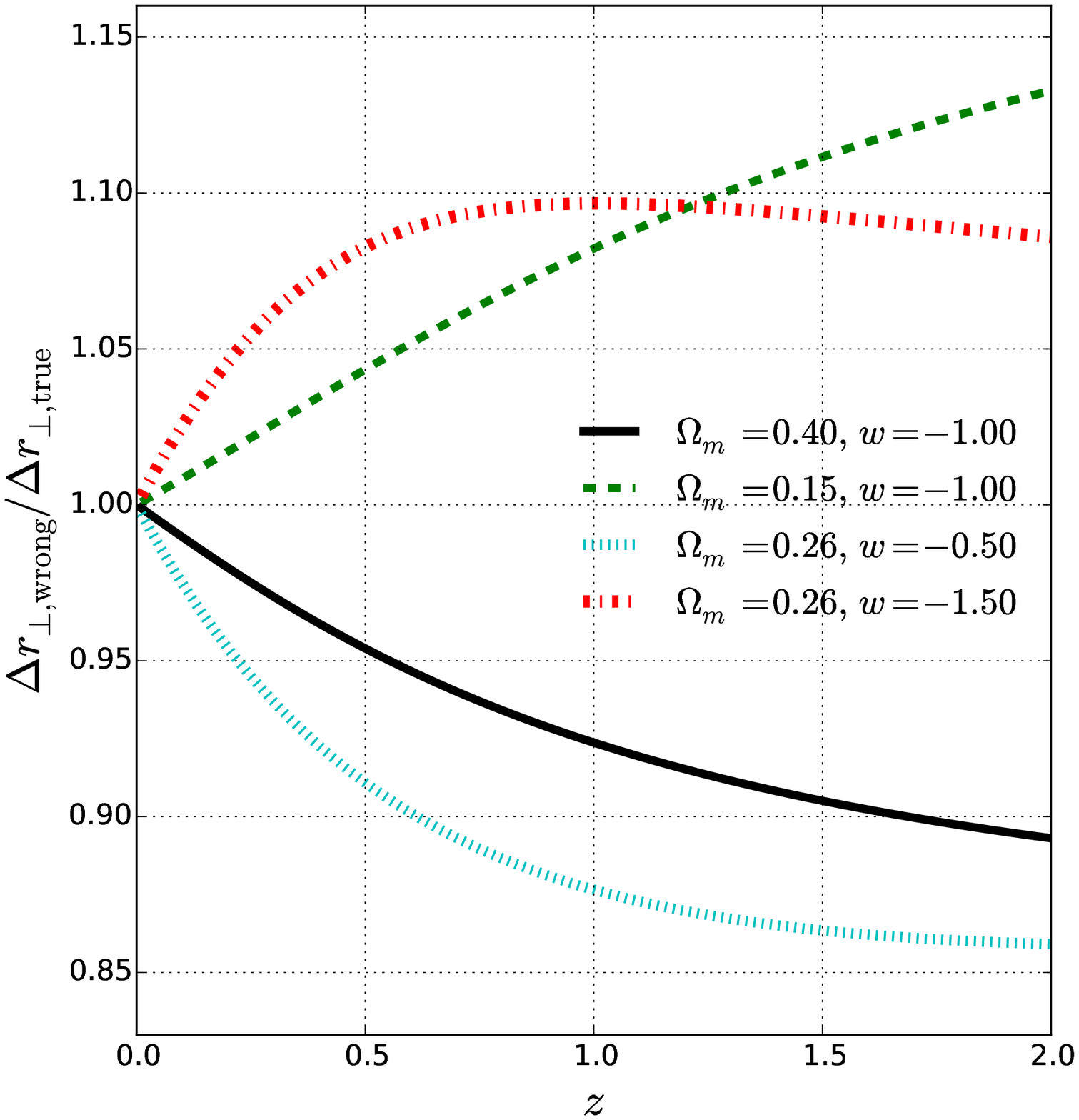}
   }
   \caption{\label{fig_xyquan}
   An illustrative example of how incorrectly assumed cosmologies can distort a distribution of points,
   assuming $\Omega_m=0.26,w=-1$ is the true cosmology. 
   Left panel shows a series of five regular squares, measured by an observer at the origin.
   Their true positions and shapes are plotted as blue dashed lines.
   When the observer adopts incorrect cosmologies when computing distances from redshifts and inferring the positions and shapes of the squares,
   they obtain distorted shapes (red solid lines).
   Right panel shows the redshift evolution of the wrongly estimated angular diameter distance (divided by the correct value).
   }
\end{figure*}

In this section we briefly discuss the scaling effect in the galaxy clustering statistics caused by assuming incorrect cosmological parameters. A more detailed description can be found in \cite{Li2014,Li2015,Li2016}.

Suppose that we are probing the size of some objects in the Universe.
We measure their redshift span $\Delta z$ and angular size $\Delta \theta$,
then compute their sizes in the radial and transverse directions using the following formulas
\begin{equation}\label{eq:distance}
\Delta r_{\parallel} = \frac{c}{H(z)}\Delta z,\ \ \Delta r_{\perp}=(1+z)D_A(z)\Delta \theta,
\end{equation}
where $H$ is the Hubble parameter and $D_A$ is the angular diameter distance.
In the particular case of a flat universe composed of a cold dark matter component and a constant EoS dark energy, they take the forms
\begin{eqnarray}\label{eq:HDA}
& &H(z) = H_0\sqrt{\Omega_ma^{-3}+(1-\Omega_m)a^{-3(1+w)}},\nonumber\\
& &D_A(z) = \frac{1}{1+z}r(z)=\frac{1}{1+z}\int_0^z \frac{dz^\prime}{H(z^\prime)},
\end{eqnarray}
where $a=1/(1+z)$ is the cosmic scale factor,
$H_0$ is the present value of Hubble parameter and $r(z)$ is the comoving distance.

When incorrect values of $\Omega_m$ and $w$ are adopted, 
the inferred $\Delta r_{\parallel}$ and $\Delta r_{\perp}$ are also incorrect,
resulting in erroneous estimation of the object's shape (AP effect) and size (volume effect).
These effects and their cosmological consequences have been studied in \cite{Li2014,Li2015,Li2016}.

In this paper we focus on the mis-estimation of the angular size $\Delta r_{\perp}$. 
The ratio of mis-estimation is
\begin{equation}\label{eq:DA}
 \alpha_{\perp} \equiv \frac{\Delta r_{\perp, \rm wrong}}{\Delta r_{\perp, \rm true}}
 = \frac{D_{A, \rm wrong}}{D_{A, \rm true}},
\end{equation}
where ``true'' and ``wrong'' denote the values in the true cosmology and wrongly assumed cosmologies respectively.

An illustration is provided in the left panels of Figure \ref{fig_xyquan}.
Suppose that the true cosmology is a flat $\Lambda$CDM with present matter ratio $\Omega_m=0.26$
and standard dark energy EoS $w=-1$.
If we were to distribute a series of perfect squares at distances ranging from 500 Mpc/h to 3\,000 Mpc/h,
and an observer located at the origin were to measure their redshifts and infer the sizes of the squares
using the distance-redshift relations of four incorrect cosmologies
\begin{eqnarray}
 & &{\rm (i)}.\ \Omega_m=0.40,\ w=-1.0, \nonumber \\ 
 & &{\rm (ii)}.\ \Omega_m=0.15,\ w=-1.0, \nonumber\\ \noindent
 & &{\rm (iii)}.\ \Omega_m=0.26,\ w=-0.5,\nonumber\\ \noindent
 & &{\rm (iv)}.\ \Omega_m=0.26,\ w=-1.5,  \nonumber \noindent 
\end{eqnarray}
then the shapes of the squares would appear distorted (AP effect),
and their sizes would be wrongly estimated (volume effect).
Cosmological models (i,iii) result in compressed volume elements (in both the angular and LOS directions),
and the degree of compression increases with increasing distance;
the situation is opposite for the other two cosmologies.

The mis-estimation of angular size, ${\Delta r_{\perp, \rm wrong}}/{\Delta r_{\perp, \rm true}}$, 
is displayed in the right panel of Figure \ref{fig_xyquan}.
In all cosmologies, ${\Delta r_{\perp, \rm wrong}}/{\Delta r_{\perp, \rm true}}$ evolves significantly in the redshift range $0\leq z\leq 2$.
As an example, when adopting the quintessence cosmology $\Omega_m=0.26,\ w=-0.5$,
the angular size is underestimated by 
8.9\%, 12.3\%, 13.6\%, 14.1\% 
at $z=0.5,1,1.5,2$.


In summary, as a consequence of incorrectly assumed cosmologies 
the size of objects is mis-estimated and the magnitude of mis-estimation depends on the redshift.
We thus use the galaxy 2pCF across the LOS to probe the mis-estimation of angular size, ${\Delta r_{\perp, \rm wrong}}/{\Delta r_{\perp, \rm true}}$.

\section{The Simulation Data}\label{sec:data}

We test the method using the mock galaxies produced by the Horizon Run 4  (HR4) N-body simulation \citep{hr4,hong2016}.

HR4 was made in a $({3.15}\ h^{-1}\rm Gpc)^3$ cube using  $6300^3$ particles with mass $m_p\simeq 9 \times 10^9 \hMsun$.
The simulation adopted the second order Lagrangian perturbation theory (2LPT) initial conditions at $z_{i}=100$
and a WMAP5 cosmology $(\Omega_{b},\Omega_{m},\Omega_\Lambda,h,\sigma_8,n_s)$  = (0.044, 0.26, 0.74, 0.72, 0.79, 0.96) \citep{komatsu 2011}.

Mock galaxies are produced from the simulation based on a modified one-to-one correspondence scheme \citep{hong2016}. 
The most bound member particles (MBPs) of simulated halos are adopted as tracers of galaxies.
The merger timescale is computed to obtain the lifetime of merged halos.
Merger trees of halos are constructed by tracking their MBPs from $z = 12$ to 0;
when a merger event occurs, the merger timescale is computed using the formula of \cite{jiang2008} to 
determine when the satellite galaxy is completely disrupted.

The resulting mock galaxies were found to well reproduce the 2pCF of the volume-limited galaxy sample from Sloan Digital Sky Survey Data Release 7 (SDSS DR7) \citep{zehavi2011}.
The mock galaxies shows a similar finger of god (FOG) feature \citep{FOG} as the observation.
The projected 2pCF of the mock and observational samples agree within 1$\sigma$ CL
on scales greater than 1 ${h^{-1}}$Mpc.

In this analysis, we use five snapshots of mock galaxies at $z=0,0.5,1,1.5,2$.
Setting a minimal halo mass of $3\times 10^{11} \hMsun$, 
we select 457, 406, 352, 306 and 228 million mock galaxies at the five redshifts,
corresponding to a number density of 
1.46, 1.30, 1.13, 0.98 and 0.73 in units of $ 10^{-2} h^{3} \rm Mpc^{-3}$
respectively.
At higher redshift, we have smaller number of galaxies, with a {\it larger bias}.
The total number of galaxies is 1.75 billion.

\begin{figure*}
   \centering{
   \includegraphics[width=16cm]{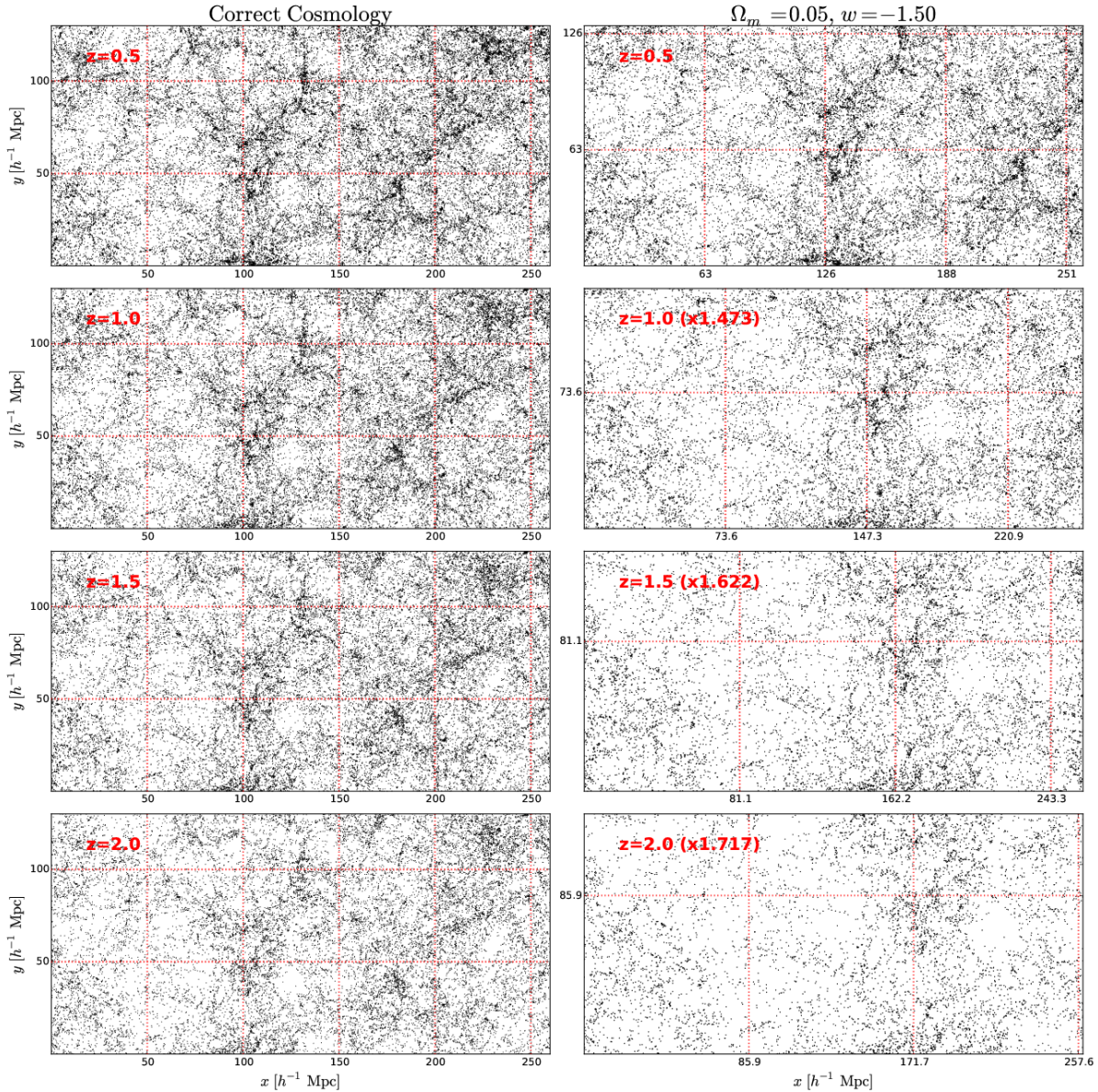}
   }
   \caption{\label{fig_scatter}
  $x$, $y$ coordinates of a $260\times130\times105 (h^{-1}{\rm Mpc})^3$ subsample of HR4 galaxies, 
  at redshifts 0.5, 1, 1.5 and 2.
  Left and right panels show the coordinates computed using the ``correct'' parameters $\Omega_m=0.26,\ w=-1$ 
  and a wrong cosmology $\Omega_m=0.05,\ w=-1.5$, respectively.
  The growth of structure strengths the clustering and make structures more compact at lower redshift.
  When the wrong cosmology is adopted, the comoving distances are upscaled by $25.6\%,47.3\%,62.2\%,71.7\%$ at redshifts of $0.5,1,1.5,2$, 
  respectively, leading to a {\it redshift evolution} in the sizes of structures.
   }
\end{figure*}

As an illustration, Figure \ref{fig_scatter} displays a subsample of galaxies at the five redshifts.
Their $x$, $y$ coordinates are computed using two sets of cosmological parameters,
the ``correct'' cosmology $\Omega_m=0.26,\ w=-1$ (left panels),
and a set of wrong cosmological parameters $\Omega_m=0.05,\ w=-1.5$ (right panels), respectively.

When the correct cosmology is adopted, the cosmic scale is correctly estimated,
and the main factor for the redshift evolution of galaxy distribution is the gravitational growth of structure.
With decreasing redshift, the clusters and filaments form and grow.
On the other hand, the wrong cosmology leads to an artificial scaling of distance.
The separations among galaxies are over-estimated by 
$25.6\%,47.3\%,62.2\%,71.7\%$ at redshifts of $0.5,1.0,1.5,2.5$,
causing a clear redshift evolution of the sizes of structures.

The growth of structure increases the clustering and results in more compact structures,
while the volume effect maintains the clustering pattern and uniformly re-scales structures on all scales.
Their imprints are different.
In the next section, we show that they affect $\xi(r_{\bot})$ measurements in different manners, 
and thus can be easily separated.






\section{Methodology}


We use the 2pCF across the LOS as a statistical tool to probe the volume effect.
The galaxy 2pCF as a function of galaxy separation perpendicular to the LOS, $\xi(r_\perp)$, is computed for snapshot data of mock galaxies at redshifts 0.5, 1, 1.5 and 2.
We adopt the Landy-Szalay estimator~\citep{1993ApJ...412...64L},
\begin{equation}
\xi(r_\perp)=\frac{DD-2DR+RR}{RR}\ ,
\end{equation}
where $DD$ is the number of galaxy--galaxy pairs, 
$DR$ the number of galaxy-random pairs, 
and $RR$ is the number of random--random pairs, 
all separated by a distance defined by $r_\perp\pm\Delta_{r_\perp}$ where we choose $\Delta_{r_\perp}=1h^{-1}\rm Mpc$.
The random catalogue consists of unclustered points uniformly distributed in the same space as the simulated data. 
In an effort to reduce the statistical variance of the estimator, we use 50 times as many random points as we have galaxies.

Considering the large number of galaxies and random points the 2pCF is computed\footnote{The code (KSTAT) for computing the correlation functions is available via
\href{https://bitbucket.org/csabiu/kstat}
{https://bitbucket.org/csabiu/kstat} and \href{http://ascl.net/code/v/1634}{http://ascl.net/code/v/1634}.
} 
part by part in subsamples with 
size of 1575 $h^{-1}{\rm Mpc}\times 1575\ h^{-1}{\rm Mpc}\times 105\ h^{-1}{\rm Mpc}$.
The `sheet'-like shape of the subsample is similar to the shape of redshift shells in the real observational case.
The Z direction with thickness $105\ h^{-1}\rm Mpc$ is treated as the radial direction (the $r$ direction) 
and the X-Y directions 
are the angular plane.
For our simulation box size we have 120 such subsamples.
The average of the measurements in all subsamples is taken as the 2pCF of the whole sample,
while the covariances of them are taken to be the covariance matrix (after multiplying by a factor of $1/\sqrt{119}$).

\subsection{Galaxy 2pCF across the LOS: at different redshifts}\label{sec_2pCF_diffz}

\begin{figure*}
   \centering{
    \includegraphics[width=2\columnwidth]{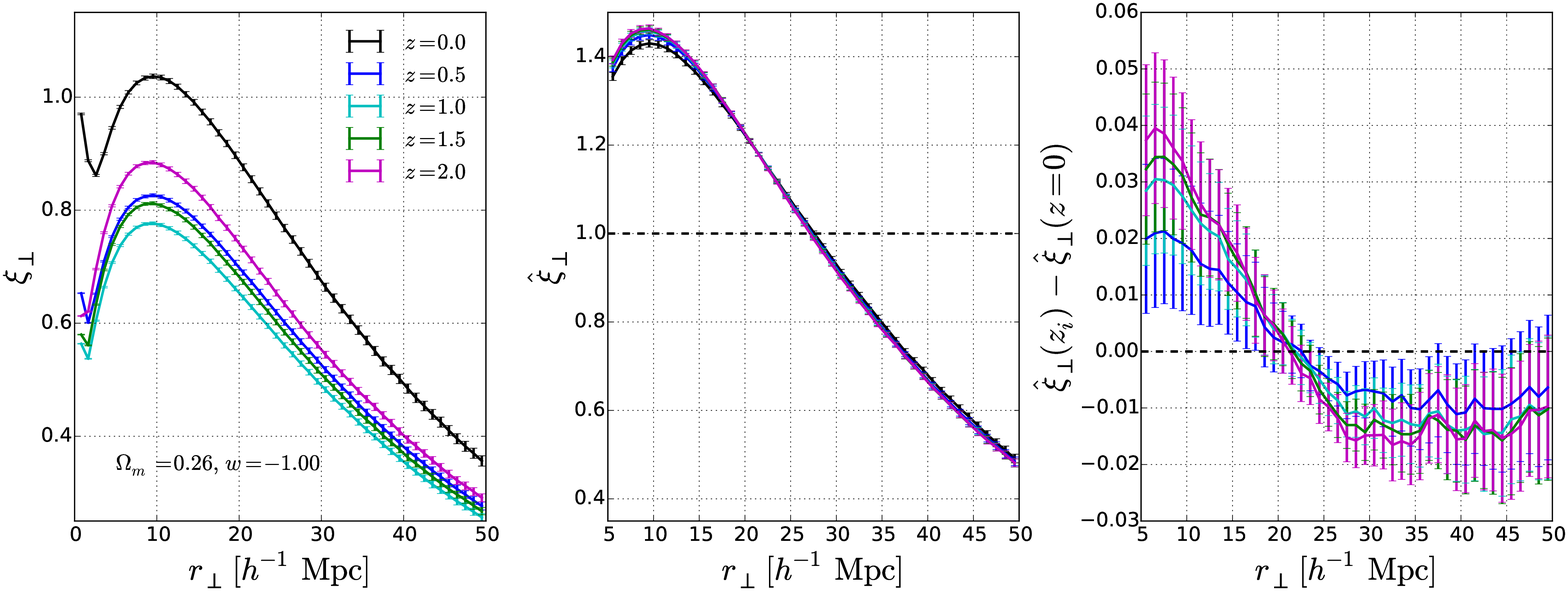}
    \includegraphics[width=2\columnwidth]{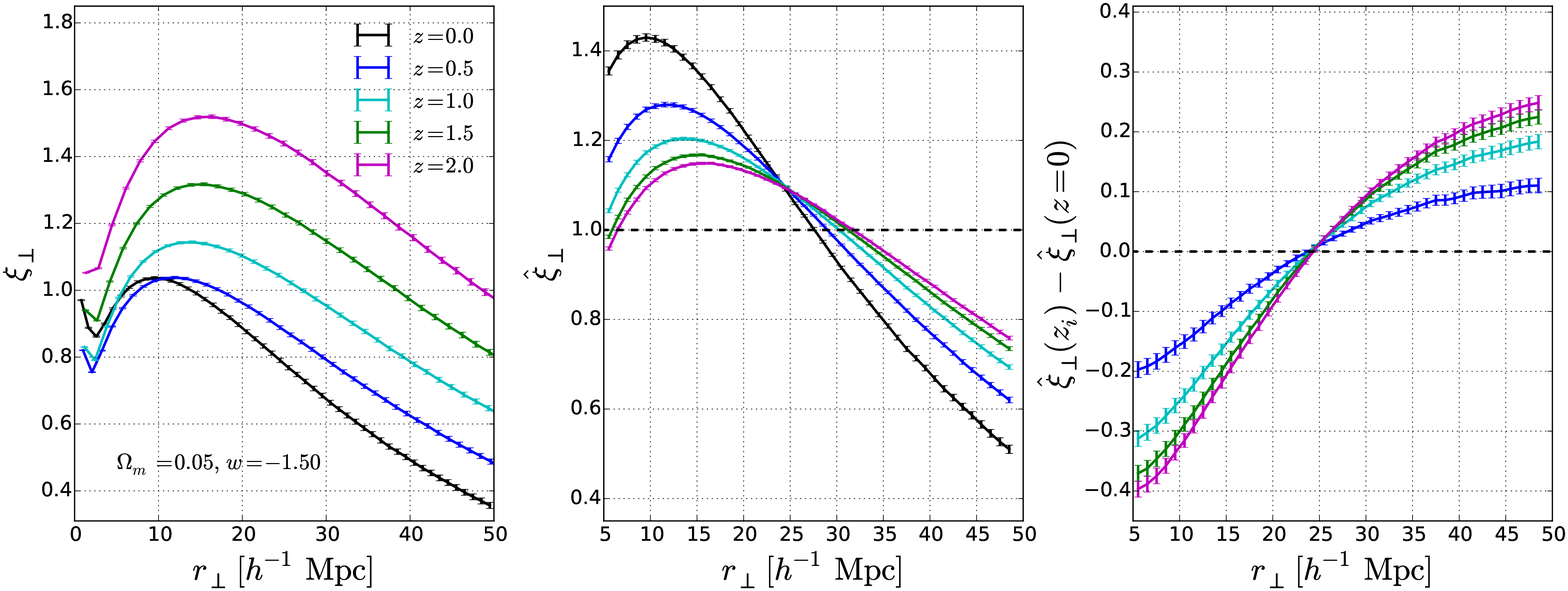}
   }
   \caption{\label{fig_diffz}
  The 2pCF across the LOS, $\xi_{r_\perp}$ (left panels), the shape of $\hat{\xi}_{r_\perp}$ (middle panels),
  and the evolution with respect to the $z=0$ value, $\hat{\xi}_{r_{\perp}}(z=z_i) - \hat{\xi}_{r_{\perp}}(z=0)$ (right panels). 
  Upper panels show that in the correct cosmology the redshift evolution of the shape of 2pCF is rather small regardless of the large redshift evolution of the amplitude, 
  whereas the lower panels show that, in a wrong cosmology $\Omega_m = 0.05,w=-1.5$, the redshift evolution of $\hat{\xi}_{r_\perp}$ is very significant.
   }
\end{figure*}

The upper-left panel of Figure \ref{fig_diffz} displays the 2pCF across the LOS measured from HR4 mock galaxies. 
We multiply $\xi$ by the separation $r_\perp$ to obtain similar statistical uncertainty on all scales.
For convenience we will use the denotation
\begin{equation}
 \xi_{r_{\perp}} \equiv r_{\perp} \xi.
\end{equation}

Among the different redshift bins there is large variation in the amplitude of $\xi(r_\perp)$.
The amplitude is proportional to the clustering strength, which is affected by the gravitational growth of structures and the galaxy bias.
The amplitude is highest at $z=0$ where the structures experienced the most growth.
$\xi(r_\perp)$ increases at $z>1$ with increasing redshift 
due to the increasing galaxy bias.

Although there is large variation in amplitude between the different redshifts bins, 
the shape of $\xi_{r_{\perp}}$  remains similar at all redshifts;
in general it peaks at $r\sim9 h^{-1}\rm Mpc$ and monotonically drops or increases at larger or smaller scales.
The only exception is the small enhancement at $r\lesssim 2 h^{-1}\rm Mpc$,
which is caused by the non-linear growth of structures 
and is much more significant at lower redshift.

In order to directly compare the shape of the 2pCF at different redshifts,
in the middle panel of Figure \ref{fig_diffz} 
we show the $r_\perp\xi$ normalized by the overall amplitude within 
$5h^{-1}{\rm Mpc} < r < 40h^{-1}{\rm Mpc}$
(here after $\hat \xi_{r_\perp}$):
\begin{equation}
 \hat \xi_{ r_\perp} \equiv  \frac{ {r_\perp\xi}(r_{\perp})}{ \int_{r_{\perp, \rm min}}^{r_{\perp, \rm max}}r_{\perp} \xi(r_{\perp}) d r_{\perp} / (r_{\perp,\rm max}-r_{\perp,\rm min}) },
\end{equation}
where we choose $r_{\perp, \rm min}=5 h^{-1}$ Mpc, $r_{\perp, \rm max}=40 h^{-1}$ Mpc in this analysis.
Below $5 h^{-1}$ Mpc the non-linear growth of structure 
leads to systematic redshift evolution which could be difficult to be reliably accounted;
On scales larger than $40 h^{-1}$ Mpc,
the analytical modeling of the shape of $\xi_{r_{\perp}}$ is relatively well understood;
one can just fit the 2pCF with the theoretical predictions \citep{BCGS2001,Salvador2014,Salvador2016} 
instead of using our method 
(although our method should also be applicable).

In the upper-middle panel, the overlapping of $\hat\xi_{r_{\perp}}$
clearly shows that there is minimal redshift evolution of the shape.
In the upper-right panel, 
we further show the residual evolution at high redshifts 
with respect to $z=0$.
There is only 1-4\% enhancement 
at $r<10h^{-1}\rm Mpc$,
and $<1.5\%$ suppression at $r>25 h^{-1}\rm Mpc$.
The trend is monotonic with redshift.

There is one detail in the analysis that should be emphasized here.
In real observations we may be selecting {\it different} types of galaxies at different redshifts.
Considering this fact in the comparison of $\hat\xi_{r_\perp}$ 
we compare subsamples of galaxies at {\it different} locations.
As an example, if at $z=0$ we take $\hat\xi_{r_\perp}$ measured within $0h^{-1}{\rm Mpc}<Z<105 h^{-1}{\rm Mpc}$,
at higher redshifts we then adopt measurement within $105h^{-1}{\rm Mpc}<Z<210 h^{-1}{\rm Mpc}$ for a comparison.
If we simply compare the 2pCF of the {\it same} subsample of galaxies at different redshifts, 
one would significantly underestimate the statistical uncertainty of $\delta \hat{\xi}_{r_\perp}$ by ignoring cosmic variance.



\subsection{Galaxy 2pCF across the LOS: cosmological effect }

The 2pCF across the LOS for galaxy positions constructed in a wrongly assumed cosmology $\Omega_m=0.05,\ w=-1.5$
is displayed in the lower panels of Figure \ref{fig_diffz}.
The scaling uniformly shifts the clustering pattern on all scales, 
leading to a biased $\hat \xi_{r_\perp}$, which is 
related with the true $\hat \xi$ as
\begin{equation}
 \hat \xi_{r_\perp,\rm wrong}(r) = \hat \xi_{r_\perp, \rm correct}(\alpha_{\perp} r),
\end{equation}
a simple consequence of the fact that the clustering pattern at scale $r$ is rescaled to $\alpha_{\perp} r$.

The redshift evolution of $\alpha_{\perp}$ leads to redshift evolution of 
$\hat \xi_{r_\perp, \rm correct}$,
which is displayed in the lower-middle panel.
The upscaling of comoving distance leads to a stretched shape.
When increasing redshift,
the peak location is shifted from $\sim9h^{-1} \rm Mpc$ to $\sim15 h^{-1}\rm Mpc$ at $z=2$,
as a result of the fact that the $\hat\xi_{r_{\perp}}$ separation is upscaled by $71.7\%$.
Correspondingly, the right panel shows that, 
compared with $\hat \xi_{r_\perp}$ at $z=0$ there is a 20-40\% change at higher redshifts.

This scaling not only changes the shape of the 2pCF but also changes the value of $ \xi_{r_\perp}$. 
As is shown in the left panel, due to the stretch of scale the amplitude is enhanced at higher redshifts;
the higher the redshift, the greater the enhancement
\footnote{The value of $\xi$ is not affected by the mis-scaling. 
But since we are using $r_\perp\xi$ rather than $\xi$, the y-axis values are affected. 
For a 20\% up-rescaling, the peak value of $r_\perp\xi$ is also increased by 20\%.}.

Although the change of amplitude could be a more significant cosmological consequence 
than the alteration of shape,
it is mixed with the other effects; i.e.
the growth of structure and a larger galaxy bias can also lead to a stronger clustering and thus an enhanced amplitude.
In order to reliably extract the cosmological information, 
we just utilize the redshift evolution of the 2pCF shape, 
which is less affected by these complicated factors.

\subsection{Systematic effects}

\begin{figure*}
   \centering{
   \includegraphics[width=1.7\columnwidth]{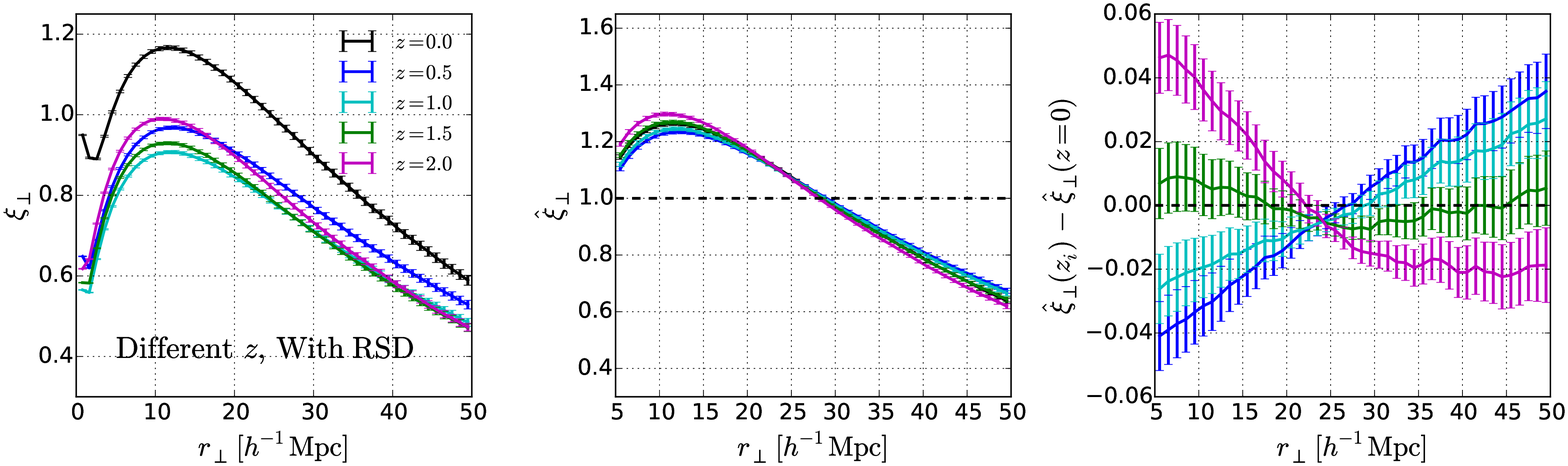}
   \includegraphics[width=1.7\columnwidth]{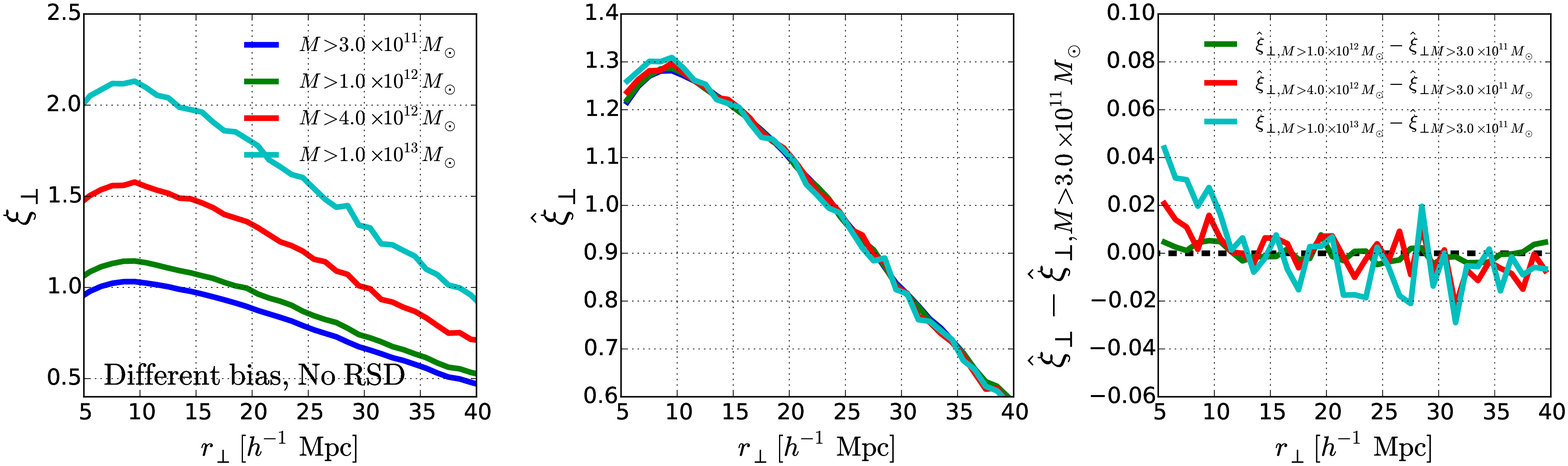}
   \includegraphics[width=1.7\columnwidth]{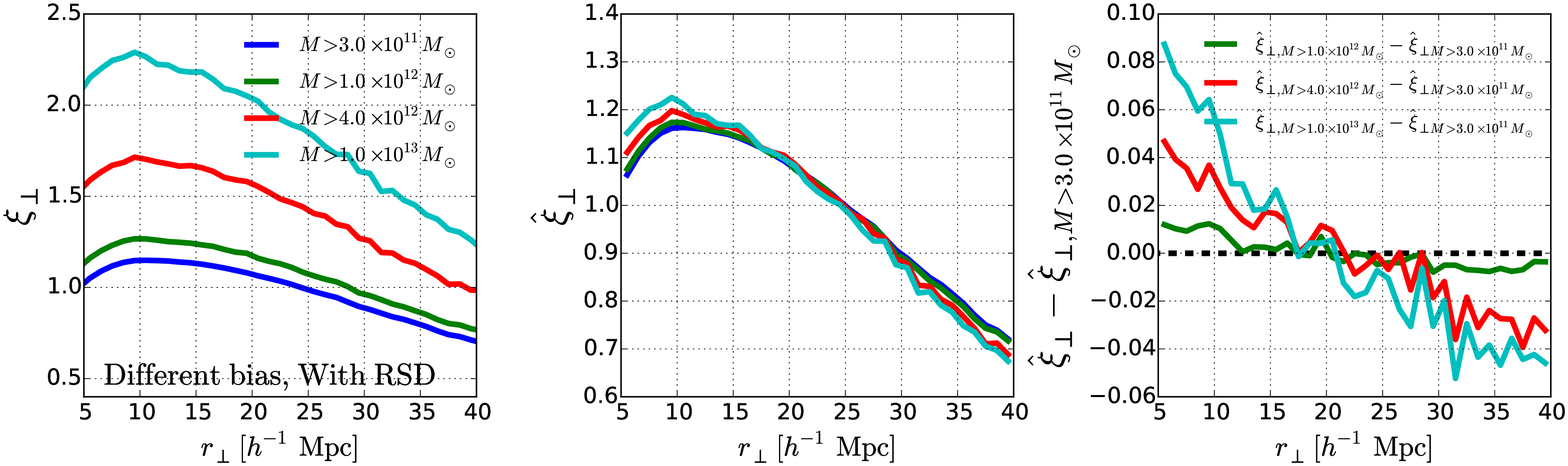}
   \includegraphics[width=1.7\columnwidth]{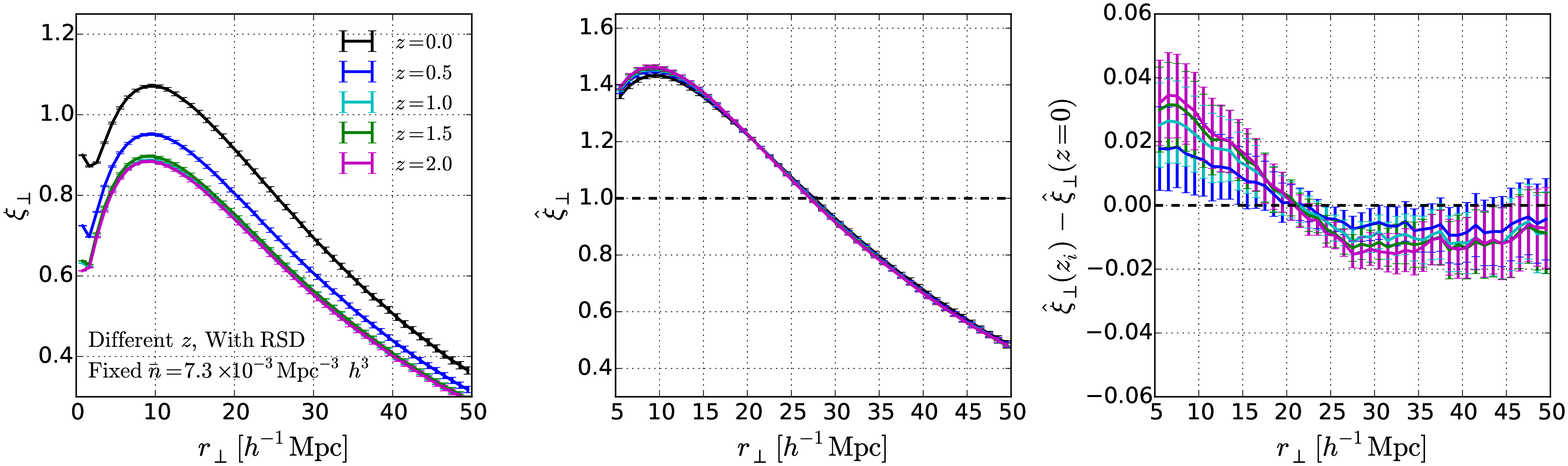}
   \includegraphics[width=1.9\columnwidth]{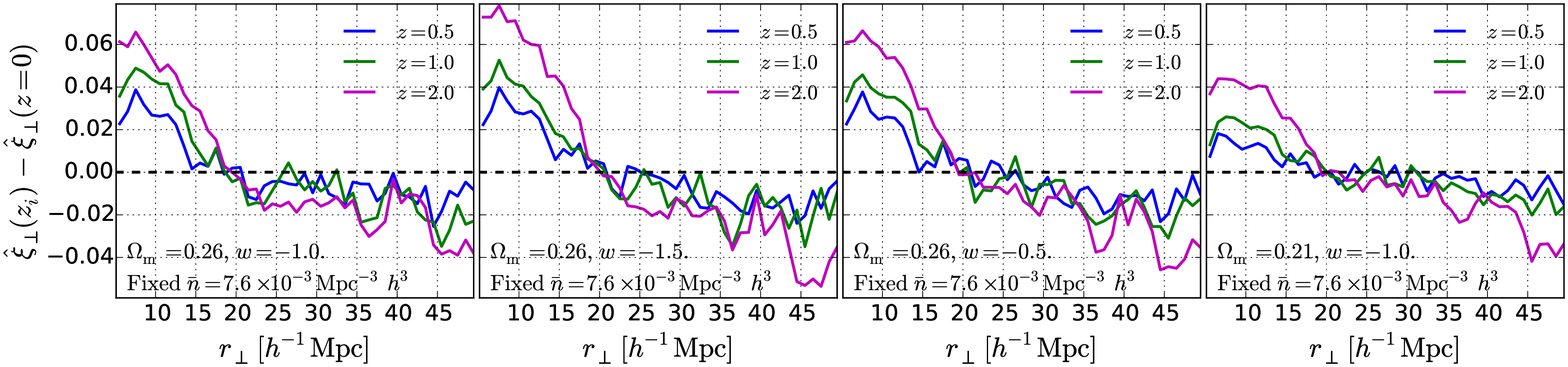}
   }
   \caption{\label{fig_sys}
  First row: $\xi_{r_\perp}$ (left), $\hat{\xi}_{r_\perp}$ (middle), and $\hat\xi_{r_\perp}(z=z_i)-\hat\xi_{r_\perp}(z=0)$ (right) calculated with the RSD effect included.
  Second row: $\xi_{r_\perp}$ (left) and $\hat{\xi}_{r_\perp}$ (middle) are shown for four different halo-mass-cuts, $3\times 10^{11},~1\times 10^{12},~4\times 10^{12},~\&~1\times 10^{13}~M_\odot$, 
  below which we remove from 2pCF calculation. 
  The right panel shows the difference in $\hat{\xi}_{r_\perp}$ between the $3\times 10^{11}~M_\odot$ mass-cut case and the other mass-cut cases. 
  Third row: The same as the second row panels, except that the RSD effect is considered.
  Fourth row: The redshift evolution in case of using {\it samples with constant number density} $\bar n=7.3\times 10^{-3} {\rm Mpc}^{-3}h^3$, 
  and the RSD effect is considered. 
  Fifth row:  The cosmological dependence of systematic effect. 
  We plot $\hat\xi_{r_\perp}(z=z_i)-\hat\xi_{r_\perp}(z=0)$ measured in four different cosmologies.
  The difference among them is $\lesssim2\%$ on all clustering scales and $\lesssim0.5\%$ on scales larger than 20 $h^{-1}\rm Mpc$.
  }
\end{figure*}

In Sec. \ref{sec_2pCF_diffz} we showed that $\hat \xi_{r_\perp}$ measured from the
constant mass cut samples at different redshifts show good agreement.
The gravitational growth of structure will not have a large impact on $\hat \xi_{r_\perp}$ on scales $\gtrsim5 h^{-1}$ Mpc,
yet there are many other factors that may induce redshift evolution
including galaxy bias, RSD, redshift error, and 
the redshift evolution of galaxies properties 
such as mass, morphology, color, concentration. 
Here we test two ``major'' systematical effects;  the RSD effect and galaxy bias.



\subsubsection{Redshift Space Distortion}

\begin{figure*}
   \centering{
    \includegraphics[width=2\columnwidth]{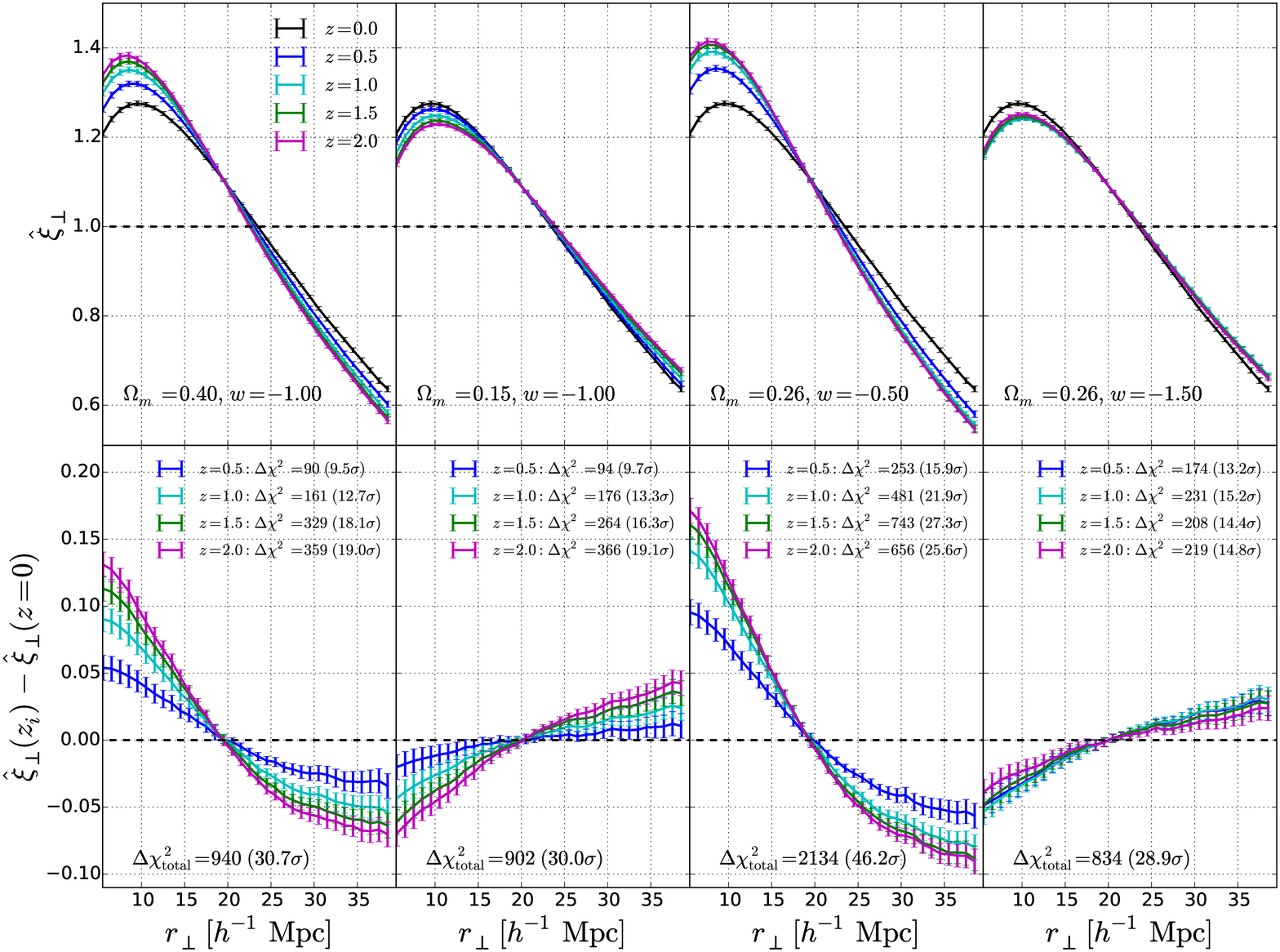}
   }
   \caption{\label{fig_cosmo}
    $\hat{\xi}_{r_\perp}$ (upper panels) and $\hat{\xi}_{r_\perp}(z=z_i) - \hat{\xi}_{r_\perp}(z=0)$ (lower panels) at several redshifts,
    for four wrong cosmologies.
    The redshift evolution of the $\hat\xi_{r_{\perp}}$ shape in these wrong cosmologies are detected at high CL.
   }
\end{figure*}

The galaxy peculiar velocity contaminates the observed redshift and distorts 
the inferred galaxy radial position.
It was the major systematic in our previous works \citep{Li2014,Li2015,Li2016}
of statistical analysis on the 3D galaxy distribution.

The effect of RSD is much milder in this work. 
The angular positions of galaxies are not shifted by RSD at all;
the only effect of RSD enter in the redshift distribution of the galaxies in our predefined redshift shell.
The galaxies observed in a survey are split into shells of subsamples, 
with different redshift ranges. This allows us to obtain 2pCF measurement at various redshifts.
RSD distorts the galaxy redshift and as a result some galaxies 
(especially those close to the boundaries of shells) are assigned to the wrong redshift shells.

We shift the radial coordinates of galaxies according to the relation 
\begin{equation}\label{eq:zvpeu}
\Delta z = (1+z) \frac{v_{{\rm Z}}}{c},
\end{equation}
where $v_{{\rm Z}}$ (Z is the third coordinate of the galaxy in the box, treated as the radial direction in this analysis) 
is the galaxy peculiar velocity in the LOS direction.
This lead to {\it misclassification} of some galaxies when we split the box into slices.

The upper panel of Figure \ref{fig_sys} shows how it affects on the 2pCF.
Comparing this plot with Figure \ref{fig_diffz} one can clearly see the effect of RSD.
The amplitude of the measured 2pCF is enhanced by $\sim 10\%$ when considering the RSD effect 
and the slop of $\hat\xi_{r_\perp}$ is suppressed.
However, the redshift evolution of $\hat\xi_{r_\perp}$ remains small,  $\lesssim4\%$ at $0.5\leq z \leq 2$.
Therefore RSD should not significantly affect the application of our method, as long as its redshift evolution is small;
the small effect of RSD can be precisely modelled by simulations and corrected.

Similar to RSD, redshift errors of galaxies
also results in fuzzy boundaries of redshift shells.
Thankfully, there are some methods for mitigating this effect in photometric surveys using, for example, pair-center binning \citep{2010MNRAS.407..520N, 2011MNRAS.415.2193R}. 

\subsubsection{Galaxy Bias}

\begin{figure*}
   \centering{
   \includegraphics[width=8cm]{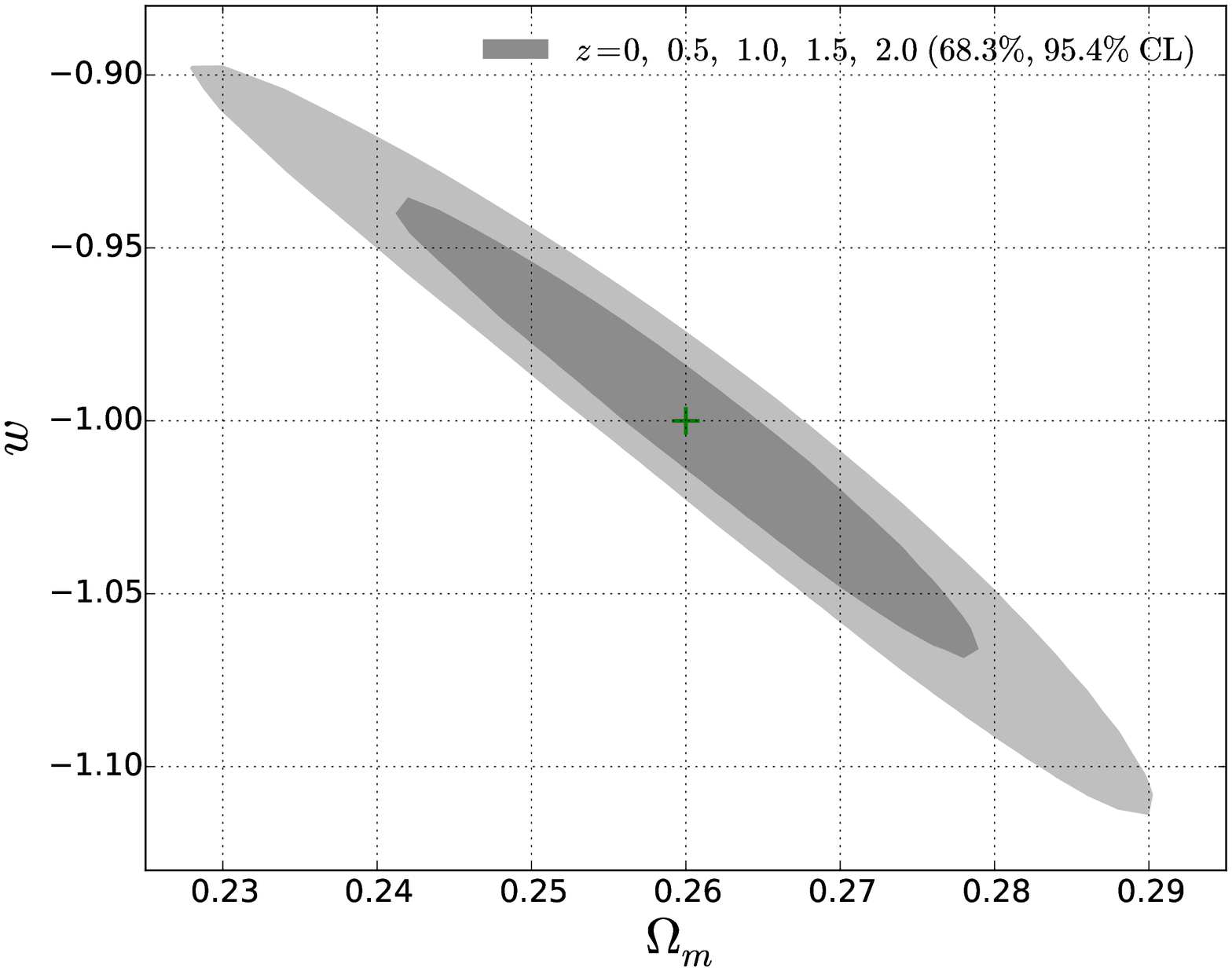}
   \includegraphics[width=8cm]{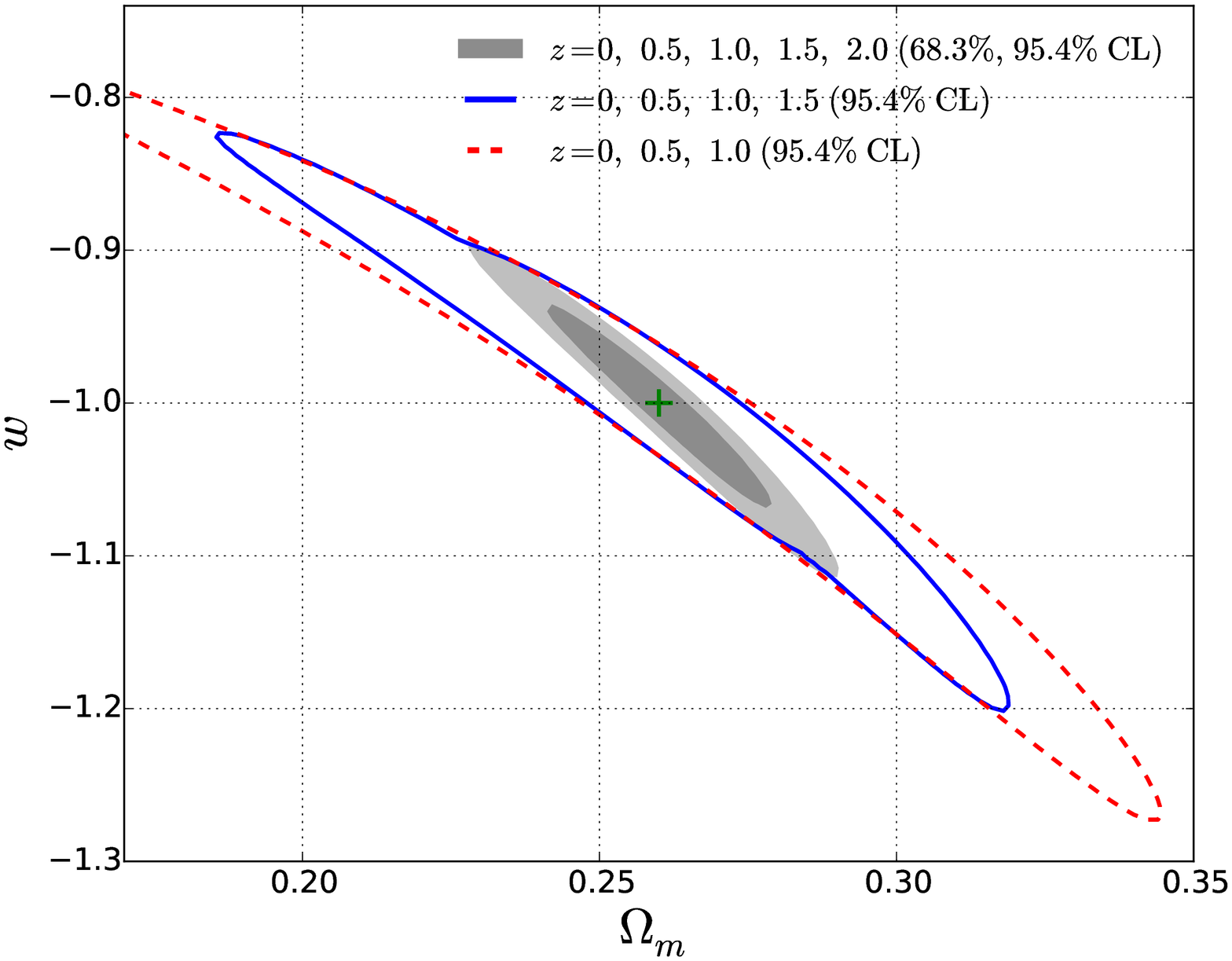}
   \includegraphics[width=8cm]{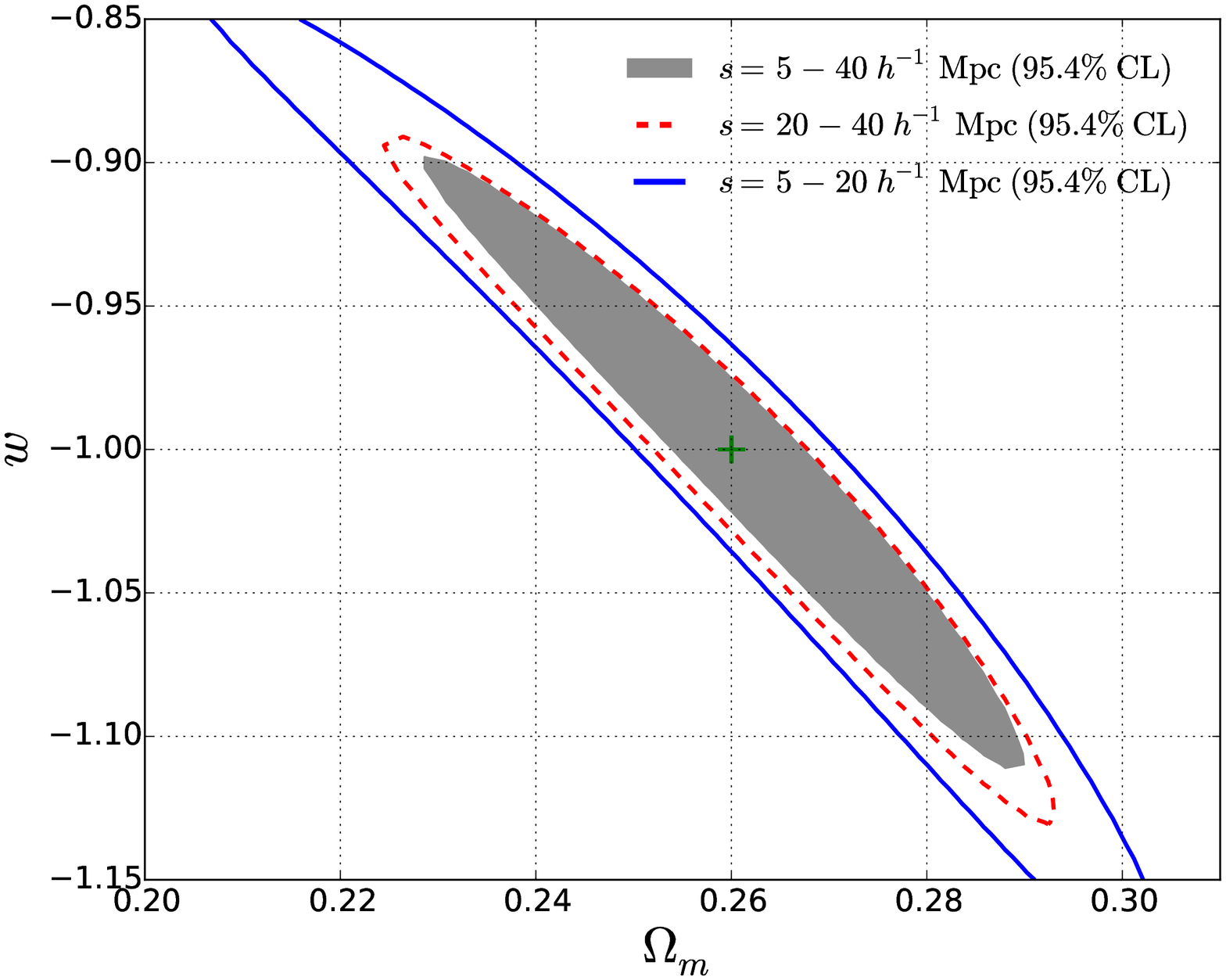}
   \includegraphics[width=8cm]{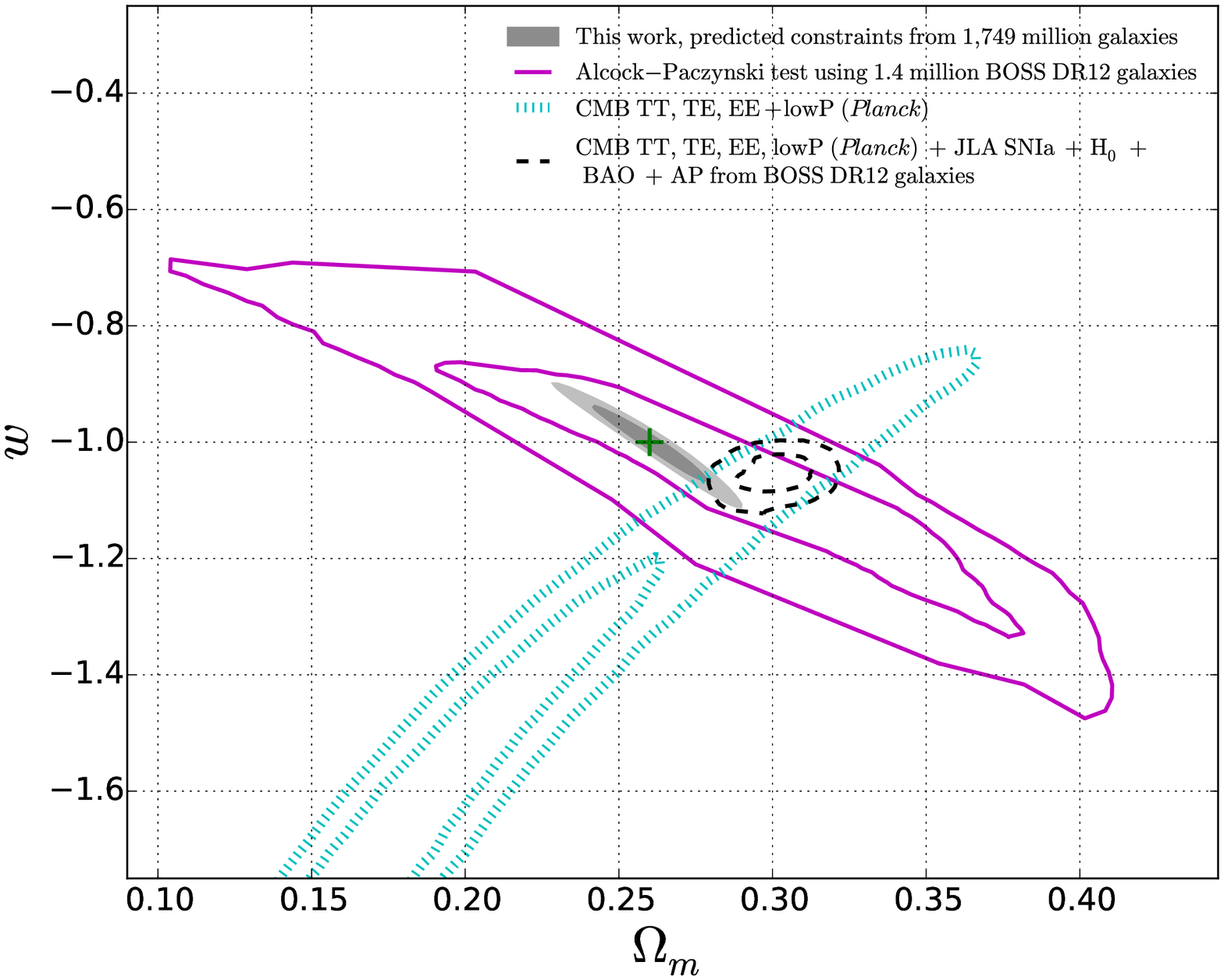}
   }
   \caption{\label{fig_contours}
   Upper left: Likelihood contours (68.3\%, 95.4\%) in the $\Omega_m-w$ plane from our method, 
   based on the redshift evolution of the angular 2pCF shape $\hat\omega_{r_\perp}$.
   The green point denotes our fiducial cosmological model and the contours denote the statistical error.
   Utilizing the HR4 snapshots data (containing hundreds of millions of galaxies at redshifts 0, 0.5, 1, 1.5, 2),
   the method lead to very tight cosmological constraint.
   Upper right: The constraints in case that one or two high redshift bins are not used.
   Lower left: The 95.4\% likelihood contours using different clustering scales.
   Lower right: A comparison of the constraint with the cosmological constraints from the current observational data.
   }
\end{figure*}

Galaxies are biased tracers of the dark matter field;
more massive galaxies reside in regions with higher density contrast
and exhibit stronger clustering (i.e. larger bias).

We vary the mass cut and check the effect of galaxy bias on $\hat \xi_{r_\perp}$.
The second row of Figure \ref{fig_sys} displays the 2pCF measured from galaxies within a $3150\times3150\times105$ $(h^{-1}\rm Mpc)^3$ volume, taken from the $z=0$ snapshot.
Four values of minimal mass limits, $3\times 10^{11} M_{\odot}$, $1\times 10^{12} M_{\odot}$, $4\times 10^{12} M_{\odot}$ and $1\times 10^{13} M_{\odot}$,
are imposed to create samples with different galaxy biases.
The measured 2pCF across the LOS are displayed.

Different mass cuts result in large variation in the amplitude of the 2pCF (left panel);
but the shape of the normalised 2pCF remains less affected (middle panel).
Compared with the sample of $M>3\times 10^{11} M_{\odot}$,
samples with mass limits of $1\times 10^{12} M_{\odot}$, $4\times 10^{12} M_{\odot}$ and $1\times 10^{13} M_{\odot}$
have amplitude of 2pCFs enhanced by $10\%,\ 50\%$ and $100\%$, 
while the change in the shape is only $0.5\%,\ 2\%$ and $4\%$, respectively.
The effect of galaxy bias becomes much less significant by utilizing the shape of the 2pCF rather than the amplitude.

The combined effect of galaxy bias and RSD results in larger systematics.
Galaxies that are more biased predominantly reside within nonlinear, over-dense regions and thus have larger larger peculiar velocities.
The third row of Figure \ref{fig_sys} shows that, 
the difference of $\hat \xi_{r_\perp}$ among different mass cut samples reaches $1.5\%,\ 4\%$ and $8\%$ in the case when including the RSD effect.

Finally, the fourth row of the Figure displays the 
redshift evolution in case of using samples with constant number density $\bar n=7.3\times 10^{-3} {\rm Mpc}^{-3}h^3$, 
with the RSD effect considered. 
We see the redshift evolution of $\hat \xi_{r_\perp}$ is as small as $\lesssim3\%$.





\subsubsection{Cosmological Dependence}
We remove systematics in our analysis using simulations with a fixed cosmology, as discussed in \S\ref{sec:data}. 
However, the systematic variation arises from various effects such as non-linear clustering and RSD, thus it should exhibit a cosmological dependence. So far we have assumed that any cosmological dependence will be small and have a minor  impact upon our results. 

In an attempt to clarify this issue, we measure the systematic effect in four different cosmologies;
the fiducial $\Omega_m=0.26$ $\Lambda$CDM cosmology,
two cosmologies with $w$ equal to 0.5 and -1.5,
and one cosmology with $\Omega_m$=0.21. These parameter choices are sufficiently deviated from the current best fit models, to ensure a good test of systematic variation.

We run four sets of simulations with these values of cosmological parameters.
Each simulation has a box size of $({1.024}\ h^{-1}\rm Gpc)^3$,
and the other settings same to the Horizon Run 4 simulation.
All simulations adopt the same random seed for the generation of initial condition,
so the only difference among them is the difference in cosmological parameters.

The fifth row of Figure \ref{fig_sys} shows the redshift evolution of $\hat \xi_{r_\perp}$ in these four different cosmologies.
Regardless of the very different cosmological parameters,
the difference among the measurements is $\lesssim2\%$ on all clustering scales,
and $\lesssim0.5\%$ on scales larger than 20 $\ h^{-1}\rm Mpc$.
So the cosmological dependence of systematic effect is not a serious problem.

A more precise approach would involve estimating the systematic effect from a set of simulations covering the relevant volume of the parameter space. 
This would remove the remaining uncertainty associated with the cosmological dependence. 
This approach would also enable the estimation of the $systematic\ error$, 
which could be important when the sample size is large enough and the systematic error becomes as small as the $statistical\ error$.


\subsubsection{More discussion}

In summary, we aim to minimise the redshift evolution of the shape of the normalised  2-point correlation function across the LOS. 
We find that this shape change is dramatic when considering a cosmological model significantly different from the fiducial model.
However even in the case of the correct cosmology there remains a small signal, due to growth of structures, RSD and galaxy bias. These systematics are removed with the aid of simulations.
The procedure of the correction of systematics is similar to what was implemented in our previous papers \cite{Li2014,Li2015,Li2016}.


The third row of Figure \ref{fig_sys} shows that, 
if having a large difference in galaxy bias between the samples,
there is a large systematic variation on small scales. 
This means that accurate modelling of systematics from simulations is required. 
In an effort to mitigate this problem, one can conduct the analysis using a volume-limited galaxy sample, 
where the redshift evolution of galaxy properties is small and the systematic effect from RSD and galaxy bias is also small,
as shown by the fourth row of Figure \ref{fig_sys}.

Another solution, to avoid the large systematics in the nonlinear regime, 
would be to increase the scale of analysis to larger separations where the non-linear effect becomes subdominant. 
Figure \ref{fig_sys} shows that, 
on scales of $s=20-50 h^{-1} {\rm Mpc}$ the systematic effect is 2-3 times smaller than that of $s=5-20 h^{-1} {\rm Mpc}$. 
Actually, we tested and found that, using $s = 5-40 h^{-1} {\rm Mpc}$ and $20-40 h^{-1} {\rm Mpc}$, 
yields cosmological constraints that are similar to each other (see \S 4). 

We do not consider the systematic effects in case of non-standard cosmological scenarios such as modified gravity and warm dark energy. 
These topics go beyond the scope of this preliminary work, however they are worthy future investigations.

\subsection{Likelihood Analysis}

Adopting the correct cosmological model will result in minimal evolution of the normalised 2pCF.
Thus we proceed to construct a quantitative likelihood estimator that reflects this property.
The procedure here is very similar to \cite{Li2014,Li2015,Li2016}.
We compare the high redshift $\hat\xi_{r_\perp}$ and the lowest redshift measurement,
\begin{equation}
 \delta \hat{\xi_{r_\perp}}(z_i,z_1) \equiv \hat{\xi_{r_\perp}}(z_i) - \hat{\xi_{r_\perp}}(z_1),
\end{equation}
and design the likelihood function to quantify
the redshift evolution of $\hat\xi_{r_\perp}$:
\begin{equation}\label{eq:chisq1}
\chi^2\equiv \sum_{i=2}^{n_{z}} \sum_{j_1=1}^{n_{r}} \sum_{j_2=1}^{n_{r}} {\bf p}(z_i,r_{j_1}) ({\bf Cov}_{i}^{-1})_{j_1,j_2}  {\bf p}(z_i,r_{j_2}),
\end{equation}
where $n_z$ is the number of redshifts bins, which is 5 in this analysis, 
$n_r$ is number of binning in $\hat{\xi}_{r_\perp}(r_\perp)$,
which is 35 since we have $r_\perp$ bins 
with width $1h^{-1}\rm Mpc$ in a range of $5h^{-1}{\rm Mpc}\leq r_\perp\leq 40h^{-1}{\rm Mpc}$.
${\bf p}(z_i,r_{j})$ is the redshift evolution of the correlation function shape with systematic effects subtracted
\begin{eqnarray}\label{eq:bfp}
 {\bf p}(z_i,r_{j}) \equiv&\ \delta \hat{\xi_\perp}(z_i,z_1,r_j) - \delta \hat{\xi}_{r_\perp,\rm sys}(z_i,z_1,r_j)
\end{eqnarray}
${\bf Cov}_i$ is the covariance matrix estimated from the ${\bf p}(z_i,r_{j})$ measured from 120 subsamples.
As mentioned previously, for a robust estimation of the covariance matrix, 
we always compare slices at different locations to 
include the cosmic variance.

The covariance matrix inferred from a finite number of samples
is always a biased estimate of the true matrix \citep{Hartlap}.
This can be corrected by rescaling the inverse covariance matrix as 
\begin{equation}
 {\bf Cov}^{-1}_{ij,\rm Hartlap} = \frac{N_s - n_{r}-2}{N_s-1} {\bf Cov}^{-1}_{ij},
\end{equation}
where $N_s=120$ is the number of mocks used in covariance estimation.
For this analysis the rescaling is as large as 1.43.
In the case that one has 2\,000 mocks the rescaling is less than 1.02.

Figure \ref{fig_cosmo} displays the 2pCF and the likelihood values when adopting 
the four incorrect cosmologies used in Figure \ref{fig_xyquan}.
For the cosmologies $\Omega_m=0.4,w=-1$ and $\Omega_m=0.26,w=-0.5$,
the compression shifts the clustering patterns to smaller scales
at higher redshift and make the 2pCFs steeper.
For the cosmologies $\Omega_m=0.15,w=-1$ and $\Omega_m=0.26,w=-1.5$,
the stretch of structure leads to a shallower slope of $\hat{\xi}_{r_\perp}$ at higher redshift.
In all cosmologies there is significant detection of redshift evolution.
We compute the $\chi^2$ values according to Equation \ref{eq:chisq1}, 
and find that these cosmologies are disfavored 
at $\gtrsim30\sigma$ CL.

\section{Cosmological constraint}

We constrain $\Omega_m$ and $w$ through a Bayesian analysis (\cite{Bayesian}; also see \cite{LB2002,Li2016}).
We assume that the likelihood takes the form
\begin{equation}\label{eq:like}
 \mathcal{L} \propto \exp\left[-\frac{\chi^2}{2}\right],
\end{equation}
and scan the parameter space in the $\Omega_m-w$ plane to obtain the 68.3\% and 95.4\% CL regions.
The result is displayed in the left panel Figure \ref{fig_contours}.

We obtain tight constraints on the two parameters.
The 2$\sigma$ contour lies within the region $0.23<\Omega_m<0.29$, $-1.1<w<-0.9$.
The two parameters degenerates strongly with each other,
and the area of the constrained region is quite small.
The thin shape of the contour implies that, 
when combining with other observational data with a different degeneracy directions (e.g. CMB, BAO),
a very tight combined constraint could be obtained.

In the case of fixing one parameter at its best-fit value and constraining the other,
one obtains very small 1$\sigma$ uncertainties of $\delta\Omega_m\approx0.003,\delta w\approx0.01$.

The simulation sample used in this analysis reaches $z=2$.
The constraint becomes weaker if we limit the highest redshift of the samples.
The upper right panel displays that, 
in case that we exclude the $z=2$ galaxies, 
the 2$\sigma$ uncertainties of $\delta\Omega_m\approx0.01,\\ \delta w\approx0.1$ (the other parameter fixed at its best-fit value).
The area of contour is further enlarged by $\approx 50\%$ if we only use galaxies at $z=$0, 0.5 and 1.

So far we simply follow the convention of \cite{Li2016} and adopt $5\hMpc <s< 40\hMpc$ as our analysis scale. 
As discussed in the \S 4.3.3, one may want to avoid large systematics in the non-linear regime. 
In this case one may adopt $20\hMpc<s<40\hMpc$,
which significantly reduces the systematic effects but does not degrade the power of the parameter constraints,
as shown by the lower left panel of Figure \ref{fig_contours}.

The choice of the redshift width $105 \hMpc$ is also slightly little arbitrary. 
It should be larger than the distance distortion from the RSD and redshift errors.
The optimistic scheme of clustering scales and redshift width should depend on the specific galaxy sample used in the analysis. 
One could adjust them to have a good balance between the power of cosmological constraint and the size of systematic effect.

Finally, to put the constraints of our new methodology in perspective, 
the lower right panel displays the constraints from Planck CMB (dotted blue, \cite{Planck2015}),
the AP analysis using BOSS DR12 galaxies (solid magenta, \cite{Li2016}),
and the joint constraint $\Omega_m = 0.301 \pm 0.006,\ w=-1.054 \pm 0.025$ 
obtained using the CMB+SNIa+BAO+AP (dashed black, \cite{6dFGS,MGS,Riess2011,JLA,Anderson2013,Li2016}).

The predicted constraints from this method clearly has an advantage due to the large sample size.
The total number of galaxies used in this work is 1.75 billion,
which is 1,000 times of the BOSS DR12 sample size.
So the constrained area is also $\sim 30$ times smaller.
Such a large sample size will be possible with LSST (the Large Synoptic Survey Telescope) \footnote{https://www.lsst.org/}, 
which probe tens of billions of galaxies at $z\lesssim2$.


\section{Concluding Remarks}
 
We have developed a method for constraining certain cosmological parameters by measuring the redshift evolution of the shape of the galaxy 2pCF across the LOS, $\hat \xi_{r_\perp}$.
We found that a wrongly adopted cosmology results in a redshift-dependent scaling in the constructed galaxy distribution,
which in turn leads to a redshift-dependent rescaling of $\hat \xi_{r_\perp}(r_\perp)$.
The redshift dependent effect is sensitive to cosmology while being relatively insensitive to the gravitational growth of structure,
the galaxy bias, and the RSD effect.
We tested our method on the HR4 mock galaxies having 457, 406, 352, 306 and 228 million galaxies at redshifts 0, 0.5, 1, 1.5, and 2.
Analyzing the redshift evolution of $\hat \xi_{r_\perp}(r_\perp)$ 
on scales $5  h^{-1} {\rm Mpc} \leq r_\perp \leq 40 h^{-1} {\rm Mpc}$, 
we derive tight constraints on $\Omega_m$ and $w$.

In this analysis we restrict our focus to the rescaling in the direction perpendicular to LOS.
One can also use the 2D 2pCF $\xi(s,\mu)$ 
as a function of both angular and radial scales to fully probe the scaling effect in the 3D galaxy distribution.

In our early work of \cite{Li2015}, 
we proposed to use the $\mu$-dependence of the galaxy clustering to apply an AP test. 
Tight cosmological constraints are derived from the BOSS galaxies \citep{Li2016}. 
In contrast, in this work we focus on the $s$-dependence of galaxy clustering. 
The two methods focus on different parts of the clustering information, and the constraints derived from the them should be fairly independent.
However, it is advisable to have an estimation of the covariance between them in the case that one is using both methods in combination
\footnote{An alternative way is to investigate the redshift evolution of the correlation map $\xi(s,\mu)$ as a function of both scale and direction, which captures all the information.}.

There are already works using the angular 2pCF to constrain cosmology
by directly comparing the theoretical 2pCF and the observed one \citep{Salvador2014,Salvador2016}.
Our method is simpler and just uses the fact that in the correct cosmology the shape of the 2pCF does not exhibit significant redshift evolution.
Our method is complementary to these works in that
it does not require accurate analytic modeling, 
and could be applicable on smaller scales. 


This method is designed to be applied to current and future LSS surveys,
with a particular emphasis on {\it photometric surveys}.
The galaxy 2pCF across the LOS could be less affected by the large redshift errors.
The Dark Energy Survey will probe hundreds of millions of galaxies at $z\lesssim1.5$
\footnote{https://www.darkenergysurvey.org/}, 
and LSST will probe tens of billions of galaxies at $z\lesssim2$.
We expect very tight cosmological constraints from these surveys.

Our series of works \cite{Li2014,Li2015,Li2016}, together with this analysis, 
and also \cite{topology,MS2016},
are introducing a new strategy of galaxy clustering statistics.
We utilize a statistical pattern which remains constant at all redshifts.
When incorrect cosmologies are adopted to construct the galaxy distribution, it introduces redshift-dependent geometric distortion,
and the invariance is broken.
The systematic effects can be corrected by performing an N-body simulation 
in a cosmology with values of parameters close enough to the underlying correct one.
This avoids the difficulty in the analytical modeling the galaxy clustering, galaxy bias and RSD,
and enables reliable cosmological constraints on relatively small clustering scales.
We hope these methods play an important role in deriving cosmological constraints from future LSS surveys. 
 
\section*{Acknowledgments}

We thank the Korea Institute for Advanced Study for providing computing resources (KIAS Center for Advanced Computation Linux Cluster System).
We would like to thank Stephen Appleby and Yi Zheng for kind helps and valuable comments.
CGS acknowledges support from the National Research Foundation (NRF,  \#2017R1D1A1B03034900). 
This work was partially supported by the
Supercomputing Center/Korea Institute of Science and
Technology Information with supercomputing resources
including technical support (KSC-2013-G2-003).



\begin{thebibliography}{}

\bibitem[Ade et al. (2015)]{Planck2015}
Ade, P.A.R., Aghanim, N., \& Arnaud, M., et al. arXiv:1502.01589

\bibitem[Alam et al.(2016)]{Alam2016}
Alam, S., Ata, M., \& Bailey, S., et al. 2016,
submitted to MNRAS (arXiv:1607.03155)


\bibitem[Alcock \& Paczynski(1979)]{AP1979}
Alcock, C., \& Paczynski, B. 1979, Nature, 281, 358  


\bibitem[Anderson et al.(2013)]{Anderson2013}
Anderson, L., Aubourg, \'E., \& Bailey, S. et al. 2014, MNRAS, 441, 24  
  

\bibitem[Ballinger, Peacock \& Heavens 1996]{Ballinger1996}
Ballinger, W.E., Peacock, J.A., \& Heavens, A.F. 1996, MNRAS, 282, 877  

\bibitem[Betoule et al.(2014)]{JLA}
Betoule, M., Kessler, R., \& Guy, J., et al. 2014, A\&A, 568, 32


\bibitem[Beutler et al.(2011)]{6dFGS}
Beutler, F., Blake, C., \& Colless, M., et al. 2011, MNRAS, 416, 3017

\bibitem[Beutler et al.(2013)]{Beutler2013}
Beutler, F., Saito, S., \& Seo, H.-J., et al. 2013, MNRAS, 443, 1065

\bibitem[Beutler et al.(2016)]{Beutler2016}
Beutler, F., Seo, H.-J., \& Saito, S., et al. 2016,
arXiv:1607.03150

\bibitem[Bernardeaua et al.(2002)]{BCGS2001}
Bernardeaua, F., Colombib, S., Gaztañagac, E., \& Scoccimarro, R. 
2002, Phys.Rept., 367, 1

\bibitem[Blake \& Glazebrook (2003)]{BG03}
Blake, C., \& Glazebrook, K. 2003, ApJ, 594, 665


\bibitem[Blake et al.(2011)]{Blake2011}
Blake, C., Glazebrook, K., \& Davis, T. M., 2011, MNRAS, 418, 1725  





\bibitem[Bueno Belloso et al. (2012)]{BB2012}
Bueno Belloso, A., Pettinari, G.W., Meures, N., \& Percival, W.J. 2012, Phys. Rev. D, 86, 023530




\bibitem[Christensen et al.(2001)]{Bayesian}
Christensen, N., Meyer, R., Knox, L., \& Luey, B. 2001, Class. Quant. Grav., 18, 2677


\bibitem[Chuang \& Wang(2012)]{ChuangWang2012}
Chuang, C.-H., \& Wang, Y. 2012, MNRAS, 426, 226  






\bibitem[Eisenstein et al. (1998)]{EHT1998}
Eisenstein, D.J., Hu, W., \& Tegmark, M. 1998, ApJ, 504, L57














\bibitem[Hartlap et al.(2006)]{Hartlap}
Hartlap J., Simon P. \& Schneider P. [astro-ph/0608064].


\bibitem[Hong et al.(2016)]{hong2016}
Hong, S.E., Park, C.,\&  Kim, J. 2016, ApJ, 823, 103

\bibitem[Jackson (1972)]{FOG}
Jackson, J., 1972, MNRAS, 156, 1

\bibitem[Jennings et al.(2011)]{Jennings2011}
Jennings, E., Baugh, C.M., \& Pascoli, S. 2011, MNRAS, 420, 1079  

\bibitem[Jeong et al.(2014)]{Jeong2014}
Jeong, D., Dai, L., Kamionkowski, M., \& Szalay, A.S. 2014, arXiv:1408.4648

\bibitem[Jiang et al.(2008)]{jiang2008}
Jiang, C.Y., Jing, Y. P., \& Faltenbacher, A., et al. 2008, ApJ, 675, 1095

\bibitem[Kaiser (1987)]{Kaiser1987}
Kaiser, N. 1987, MNRAS, 227, 1




\bibitem[Kim et al.(2015)]{hr4}
Kim, J., Park, C., L'Huillier, B., \& Hong, S. E. 2015, JKAS, 48, 213

\bibitem[Kim et al.(2011)]{horizonrun}
Kim, J., Park, C., Rossi, G., Lee, S.M., \& Gott, J.R. 2011, JKAS, 44, 217  


\bibitem[Nock et al.(2010)]{2010MNRAS.407..520N} Nock, K., Percival, W.~J., \& Ross, A.~J.\ 2010, \mnras, 407, 520 


\bibitem[Komatsu et al.(2011)]{komatsu 2011}
Komatsu, E., Smith, K. M., \& Dunkley, J., et al. 2011, ApJS, 192, 18  



\bibitem[Landy \& Szalay(1993)]{1993ApJ...412...64L} 
Landy, S.D., \& Szalay, A.S.\ 1993, ApJ, 412, 64 


\bibitem[Lavaux \& Wandelt(2012)]{LavausWandelt1995}
Lavaux, G., \& Wandelt, B.D. 2012, ApJ, 754, 109  


\bibitem[Lewis \& Bridle (2002)]{LB2002}
Lewis, A., \& Bridle, S. 2002, Phys. Rev. D, 66, 103511


\bibitem[Li et al.(2011)]{Li2011}
Li, M., Li, X.-D., Wang, S., \& Wang, Y. 2011, Commun. Theor. Phys., 56, 525

\bibitem[Li et al.(2014)]{Li2014}
Li, X.-D., Park, C., Forero-Romero, J., \& Kim, J. 2014, ApJ, 796, 137

\bibitem[Li et al.(2015)]{Li2015}
Li, X.-D., Park, C., Sabiu, C.G., \& Kim, J. 2015, MNRAS, 450, 807 

\bibitem[Li et al.(2016)]{Li2016}
Li, X.-D., Park, C., Sabiu, C.G., \& Kim, J. 2016, submitted to ApJ



\bibitem[Linder et al.(2014)]{Linder2013}
Linder, E.V., Minji, O., Okumura, T., Sabiu, C.G., \& Song, Y.-S. 2014, Phys. Rev. D, 89, 063525  

\bibitem[L{\'o}pez-Corredoira(2014)]{2014ApJ...781...96L} 
L{\'o}pez-Corredoira, M.\ 2014, ApJ, 781, 96 

\bibitem[Mao et al. (2016)]{Qingqing2016}
Mao, Q., Berlind, A.A., Scherrer, R.J., et al. 2016, submitted to ApJ

\bibitem[Marinoni \& Buzzi(2010)]{Marinoni2010}
Marinoni, C., \& Buzzi, A. 2010, Nature, 468, 539  

\bibitem[Matsubara \& Suto(1996)]{Matsubara1996}
Matsubara T., \& Suto, Y. 1996, ApJ, 470, L1  



\bibitem[Morandi \& Sun (2016)]{MS2016}
Morandi, A., \& Sun, M. arXiv:1601.03741


\bibitem[Outram et al.(2004)]{Outram2004}
Outram, P.J., Shanks, T., Boyle, B.J., Croom, S.M., Hoyle, F., Loaring, N.S., 
Miller, L., \& Smith, R.J. 2004, MNRAS, 348, 745  

\bibitem[Parejko et al.(2013)]{Parejko2013}
Parejko J.K., et al., 2013, MNRAS, 429, 98



\bibitem[Park \& Kim(2010)]{topology}
Park, C., \& Kim, Y.-R. 2010, ApJL, 715, L185  





\bibitem[Perlmutter et al.(1999)]{Perl1999}
Perlmutter, S., Aldering, G., \& Goldhaber, G., et al. 1999, ApJ, 517, 565  

\bibitem[Press \& Shechter(1974)]{PS1974}
Press, W.H., \& Schechter, P.L. 1974, ApJ, 187, 425

\bibitem[Reid et al.(2012)]{Reid2012}
Reid, B., Samushia, L., \& White, M., et al. 2012, MNRAS, 426, 2719  


\bibitem[Riess et al.(1998)]{Riess1998}
Riess, A.G., Filippenko, A.V., \& Challis, P., et al. 1998, AJ, 116, 1009  

\bibitem[Riess et al.(2011)]{Riess2011}
Riess, A.G., Macri, L., \& Casertano, S., et al. 2011, ApJ, 730, 119

\bibitem[Ross et al.(2011)]{2011MNRAS.415.2193R} Ross, A.~J., Percival, W.~J., Crocce, M., Cabr{\'e}, A., \& Gazta{\~n}aga, E.\ 2011, \mnras, 415, 2193 

\bibitem[Ross et al.(2015)]{MGS}
Ross, A.J., Samushia, L., \& Howlett, C., et al. 2015, MNRAS, 449, 835

\bibitem[Ross et al.(2016)]{2016MNRAS.tmp.1473R} Ross, A.~J., Beutler, F., Chuang, C.-H., et al.\ 2016, \mnras,  

\bibitem[Ryden(1995)]{Ryden1995}
Ryden, B.S. 1995, ApJ, 452, 25  



\bibitem[Sabiu \& Song(2016)]{2016arXiv160302389S} Sabiu, C.~G., \& Song, Y.-S.\ 2016, arXiv:1603.02389 

\bibitem[Salvador et al. (2014)]{Salvador2014}
Salvador, S.A., S\'{a}nchez, A.G., Padilla, N.D., \& Baugh, C.M. 2014, MNRAS, 443, 2612

\bibitem[Salvador et al. (2016)]{Salvador2016}
Salvador, S.A., S\'{a}nchez, A.G., \& Grieb, J.N., et al. 2016, submitted to MNRAS

\bibitem[S\'{a}chez et al. (2006)]{Sanchez2006}
S\'{a}chez, A.G., Baugh, C.M., \& Percival, W.J. et al. 2006, MNRAS, 366, 187

\bibitem[S\'{a}nchez et al. (2009)]{Sanchez2009}
S\'{a}nchez, A. G., Crocce, M., Cabr\'{e}, A., Baugh, C.M., \& Gaztaaga, E. 2009, MNRAS, 400, 1643

\bibitem[Sanchez et al.(2016)]{Sanchez2016}
S\'{a}nchez, A. G., Scoccimarro, R., \& Crocce, M., et al.
arXiv:1607.03147






\bibitem[Seo \& Eisenstein (2003)]{SE03}
Seo, H.-J., \& Eisenstein, D.J. 2003, ApJ, 598, 720



\bibitem[Song et al.(2014)]{2014arXiv1407.2257S} 
Song, Y.S., Sabiu, C.G., 
Okumura, T., Oh, M., \& Linder, E.V.\ 2014, JCAP, 12, 005 

\bibitem[Sutter et al.(2014)]{Sutter2014}
Sutter, P.M., Pisani, A., Wandelt, B.D., \& Weinberg, D.H. 2014, MNRAS, 443, 2983




\bibitem[Vargas-Maga{\~n}a et al.(2014)]{2014MNRAS.445....2V} Vargas-Maga{\~n}a, M., Ho, S., Xu, X., et al.\ 2014, \mnras, 445, 2 

\bibitem[Viana \& Liddle(1996)]{VL1996}
Viana, P.T.P., \& Liddle, A.R. 1996, MNRAS, 281, 323



\bibitem[Weinberg (1989)]{SW1989}
Weinberg, S. 1989, Reviews of Modern Physics, 61, 1

\bibitem[Weinberg et al. (2013)]{DHW2013}
Weinberg, D.H, Mortonson, M.J., Eisenstein, D.J., et al. 2013, Physics Reports, 530, 87



\bibitem[Yoo \& Watanabe(2012)]{2012IJMPD..2130002Y} Yoo, J., \& Watanabe, Y.\ 2012, International Journal of Modern Physics D, 21, 1230002 


\bibitem[Zehavi et al.(2011)]{zehavi2011}
Zehavi, I., Zheng, Z., \& Weinberg, D.H., et al. 2011, ApJ, 736, 59




\end{thebibliography}
\end{document}